\documentclass[twocolumn,longbib]{aastex701}
\makeatletter
\renewcommand{\frontmatter@title@above}{}
\makeatother
\usepackage{amsmath}
\usepackage{graphicx}
\usepackage{booktabs}
\usepackage{xspace}
\usepackage{tikz}
\usetikzlibrary{arrows.meta, positioning, calc}

\newcommand{\ohrc}{OHRC\xspace}

\makeatletter
\@ifclasswith{aastex701}{anonymous}{%
  \newcommand{\authorcredit}{ISRO}%
  \newcommand{\authorname}{the first author}%
  \newcommand{\suppref}{Anonymous}%
  \newcommand{\suppyear}{2026}%
  \newcommand{\ifdeanon}[2]{#2}
}{%
  \newcommand{\authorcredit}{ISRO/C.~Tungathurthi}%
  \newcommand{\authorname}{C.~Tungathurthi}%
  \newcommand{\suppref}{Tungathurthi}%
  \newcommand{\suppyear}{2026}%
  \newcommand{\ifdeanon}[2]{#1}
}
\makeatother

\newcommand{\mppx}{m\,pixel$^{-1}$}

\begin{document}

	\title{Geodetically Anchored 0.30\,m Digital Elevation Model of the Chandrayaan-3 Vikram Landing Site from Chandrayaan-2 Orbital High Resolution Camera (OHRC) Stereo Imagery}

	\author[0009-0008-2857-9700]{Chandra Tungathurthi}
	\email{c@tungathurthi.com}
	\affiliation{Independent Researcher; \href{mailto:c@tungathurthi.com}{c@tungathurthi.com}}

	\shorttitle{OHRC DEM of the Chandrayaan-3 Landing Site}
	\shortauthors{Tungathurthi}

	\begin{abstract}
		ISRO's terrain characterization and hazard mapping from Chandrayaan-2 Orbiter High Resolution Camera (OHRC) stereo imagery were central to the site selection and safe landing of Chandrayaan-3 --- the first successful landing in the lunar south polar region. However, these derived elevation products were generated with a proprietary pipeline and have not been publicly released. We present a 0.30\,m\,pixel$^{-1}$ digital elevation model (DEM) of the Chandrayaan-3 Vikram landing site produced using a fully open photogrammetric workflow based on ISIS, the Ames Stereo Pipeline, and ALE, achieving sub-meter resolution comparable to mission-reported products.

		The reconstruction covers a 2.18\,$\times$\,2.24\,km region with 91.2\% valid pixel coverage, a median triangulation error of 8.1\,cm, and an estimated relative vertical precision of 40--50\,cm. At sub-meter scale, the Vikram lander and Pragyan rover are individually resolved as discrete topographic features. Geodetic alignment to an LROC NAC stereo DEM \citep{Henriksen2016} achieves $\sim$30\,m horizontal accuracy. Pixel-wise validation against the same surface at 3\,m resolution confirms negligible vertical bias (median $\Delta z$ = $+$0.28\,m) and robust dispersion (NMAD = 2.88\,m). A key technical finding is that stable OHRC stereo convergence requires Community Sensor Model (CSM) camera models; the legacy ISIS camera model failed to produce stable solutions across two independent sites.

		At 0.30\,m, these DEMs complement the LROC NAC DTMs ($\sim$1\,m) that serve as the current standard for lunar surface operations, resolving sub-meter hazards that fall below the NAC detection threshold. Applied to the extensive OHRC south polar archive using publicly archived data, this methodology provides the broader community with an independent capability for hazard mapping, terrain characterization, and landing site analysis for upcoming missions including Chandrayaan-4, LUPEX, and Artemis.
	\end{abstract}

	\keywords{Moon --- lunar surface --- photogrammetry --- digital elevation models --- remote sensing --- planetary mapping --- topography --- image processing}

	\section{Introduction}
	\label{sec:intro}

	High-resolution topographic data underpins lunar landing site selection, hazard avoidance, traverse planning, geomorphological analysis, and post-landing reconstruction of spacecraft events.
	The spatial resolution and geodetic consistency of available elevation products directly constrain the scale at which surface processes and engineering constraints can be assessed.
	The Lunar Orbiter Laser Altimeter (LOLA) aboard NASA's Lunar Reconnaissance Orbiter provides global gridded elevation coverage at approximately 30\,m resolution, with along-track shot spacing of $\sim$57\,m and locally enhanced polar coverage approaching $\sim$5\,m \citep{Smith2010}.
	Photogrammetric digital elevation models (DEMs) derived from Lunar Reconnaissance Orbiter Camera (LROC) Narrow Angle Camera imagery achieve $\sim$2\,m resolution at selected sites \citep{Robinson2010}.
	However, sub-meter orbital topography remains limited and is typically generated through mission-internal processing pipelines.

	The Orbiter High Resolution Camera (\ohrc) aboard Chandrayaan-2 provides panchromatic pushbroom imaging (0.45--0.70\,$\mu$m) with a nominal ground sampling distance of 0.25\,\mppx from 100\,km altitude \citep{Chowdhury2020}.
	The instrument employs a 12{,}000-pixel Time Delay Integration (TDI) CCD detector with a 300\,mm primary mirror and 2046\,mm effective focal length, yielding a $\sim$3\,km nadir swath.
	OHRC was designed for stereo imaging: the spacecraft performs pitch maneuvers on consecutive orbits (approximately 5$^\circ$ and 25$^\circ$ off-nadir) to image the same terrain from distinct viewing geometries within a single orbital cycle \citep{Chowdhury2020}.
	During low-altitude operations at 63--70\,km, the ground sampling distance improves to 16--18\,cm\,pixel$^{-1}$, making OHRC the highest-resolution orbital imager currently operating at the Moon.
	Despite this capability, independent generation of geodetically consistent DEMs from OHRC stereo imagery presents nontrivial challenges.
	Stereo acquisition via pitch maneuvers, combined with the absence of reconstructed state vectors and reliance on spacecraft attitude kernels, can introduce substantial horizontal offsets and rotational misfits in unaligned products.
	ISRO's Space Applications Centre processes OHRC stereo pairs using a proprietary pipeline (OPTIMUS), achieving reported horizontal and vertical resolutions of approximately 28\,cm and 100\,cm, respectively \citep{Amitabh2021, Amitabh2023}.
	However, these derived elevation products are not publicly distributed.
	While raw imagery and navigation data are available through the PRADAN\footnote{\url{https://pradan.issdc.gov.in/ch2/}} archive, an independently reproducible and geodetically validated OHRC DEM workflow has not previously been documented.

	In this paper, we present a 0.30\,m\,pixel$^{-1}$ digital elevation model of the Chandrayaan-3 Vikram landing site (Shiv Shakti Point) derived from OHRC stereo imagery.
	The reconstruction achieves a median triangulation error of 8.1\,cm with 91.2\% valid pixel coverage over a 2.18\,$\times$\,2.24\,km region, corresponding to an estimated relative vertical precision of 40--50\,cm.
	The workflow is implemented entirely using publicly available tools, including ISIS \citep{ISIS2023}, the Ames Stereo Pipeline (ASP; \citealt{Beyer2018}), and the Abstraction Library for Ephemerides (ALE; \citealt{ALE2023}).
	We document the complete processing chain, address integration challenges specific to Chandrayaan-2 data, and demonstrate that stable stereo convergence requires Community Sensor Model (CSM) camera models rather than the legacy ISIS camera implementation.
	We further document a multi-stage geodetic alignment process in which an initial LOLA-based approach failed to constrain horizontal position, and subsequent alignment to an LROC NAC stereo DEM using manual tie-points achieved $\sim$30\,m accuracy.
	Pixel-wise validation against the LROC NAC DEM at 3\,m resolution establishes negligible vertical bias (median $\Delta z$ = $+$0.28\,m) and robust dispersion of 2.88\,m (NMAD).

	\section{Instrument and Data}
	\label{sec:data}

	\subsection{OHRC Instrument Overview}

	Table~\ref{tab:ohrc_params} summarizes the key parameters of the OHRC instrument.
	The camera uses a Ritchey--Chr\'{e}tien optical design with field-correcting optics, providing high image quality with minimal distortion over its narrow field of view ($\pm 0.86^\circ$) \citep{Chowdhury2020}.
	Multiple TDI stages allow optimized signal-to-noise under varying illumination conditions, a capability that is particularly valuable for polar observations at low solar elevation angles.

	\begin{deluxetable}{lc}
		\tablecaption{OHRC Instrument Parameters \label{tab:ohrc_params}}
		\tablehead{\colhead{Parameter} & \colhead{Value}}
		\startdata
		Spectral range & 0.45--0.70\,$\mu$m \\
		Detector & 12{,}000-pixel TDI CCD \\
		Primary mirror diameter & 300\,mm \\
		Effective focal length & 2046\,mm \\
		GSD (100\,km altitude) & 0.25\,\mppx \\
		GSD (65\,km altitude) & $\sim$0.16\,\mppx \\
		Nadir swath (100\,km) & $\sim$3\,km \\
		Field of view & $\pm 0.86^\circ$ \\
		Stereo convergence angles & $\sim$5$^\circ$, $\sim$25$^\circ$ \\
		Data format & PDS4 \\
		\enddata
	\end{deluxetable}

	As of early 2026, the Chandrayaan-2 orbiter has released over 200 OHRC images through the PRADAN archive, totaling approximately 230\,GB.
	An analysis of the complete OHRC metadata catalogue\ifdeanon{%
		\footnote{\authorname, ``Seeing in Unprecedented Detail: Key Operational Insights into Chandrayaan-2 Orbital High-Resolution Camera,'' Moon and Beyond, 2024; \url{https://moonandbeyond.blog/p/chandrayaan-2-ohrc-key-insights}}%
	}{%
		\footnote{``Seeing in Unprecedented Detail: Key Operational Insights into Chandrayaan-2 Orbital High-Resolution Camera,'' available upon request.}%
	} has identified distinct operational phases including a shift toward polar imaging in early 2024, consistent with site characterization for the upcoming LUPEX mission.

	\subsection{Stereo Pair Selection}

	We used Level-1 calibrated data products from PRADAN, which have radiometric and geometric corrections already applied \citep{Chowdhury2020}.
	Level-0 (raw) images were initially attempted but the processing was ultimately performed with Level-1 products.

	\subsubsection{The Soma Lunar Data Portal}

	Identification of viable stereo pairs from the OHRC archive is a non-trivial task: it requires cross-referencing image footprints, acquisition times, spacecraft attitude (roll, pitch, and yaw), solar geometry, and spatial overlap.
	ISRO's PRADAN portal provides browse capabilities, but at the time this work began, no location-based search tool existed for OHRC data; ISRO's own Chandrayaan-2 browse interface was released subsequently.

	To address this, we developed ``Soma,'' a custom lunar data portal (Figure~\ref{fig:soma}) that ingests the full OHRC metadata catalogue and allows interactive search by geographic location (click-to-query), acquisition date, spatial resolution, and solar illumination geometry.
	Users can identify candidate data products covering a region of interest and evaluate their suitability for stereo processing by inspecting spacecraft attitude parameters and temporal separation.
	The full OHRC positional catalogue and imagery archive ($>$120\,GiB) were downloaded and indexed for this work.

	Soma has been used in all of our OHRC analyses to date, including the identification of the IM-2 Athena lander\ifdeanon{%
		\footnote{\authorname, ``IM-2 Athena: An Almost-Successful South Polar Landing,'' Moon and Beyond, 2025; \url{https://moonandbeyond.blog/p/im-2-athena-imaged-chandrayaan-2-ohrc}}%
	}{%
		\footnote{``IM-2 Athena: An Almost-Successful South Polar Landing,'' available upon request.}%
	}, the JAXA SLIM landing reconstruction\ifdeanon{%
		\footnote{\authorname, ``SLIM: Unyielding Defiance Amidst Adversities,'' Moon and Beyond, 2024; \url{https://moonandbeyond.blog/p/how-slim-jaxa-landed-on-moon}}%
	}{%
		\footnote{``SLIM: Unyielding Defiance Amidst Adversities,'' available upon request.}%
	}, and the present DEM work.
	The portal complements existing tools such as Arizona State University's QuickMap by providing Chandrayaan-2-specific search capabilities not available elsewhere.

	\begin{figure*}
		\centering
		\includegraphics[width=0.48\textwidth]{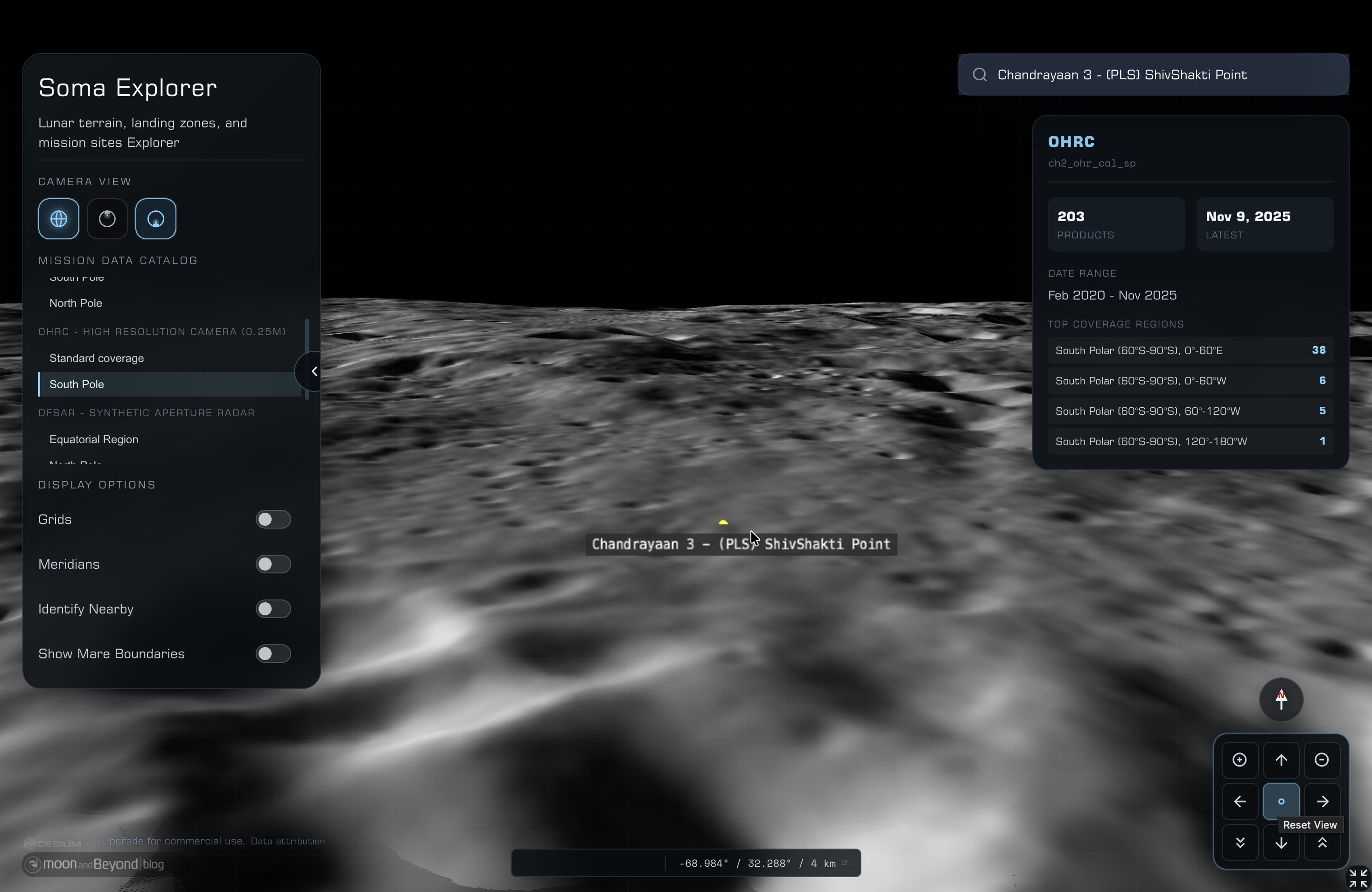}\hfill
		\includegraphics[width=0.48\textwidth]{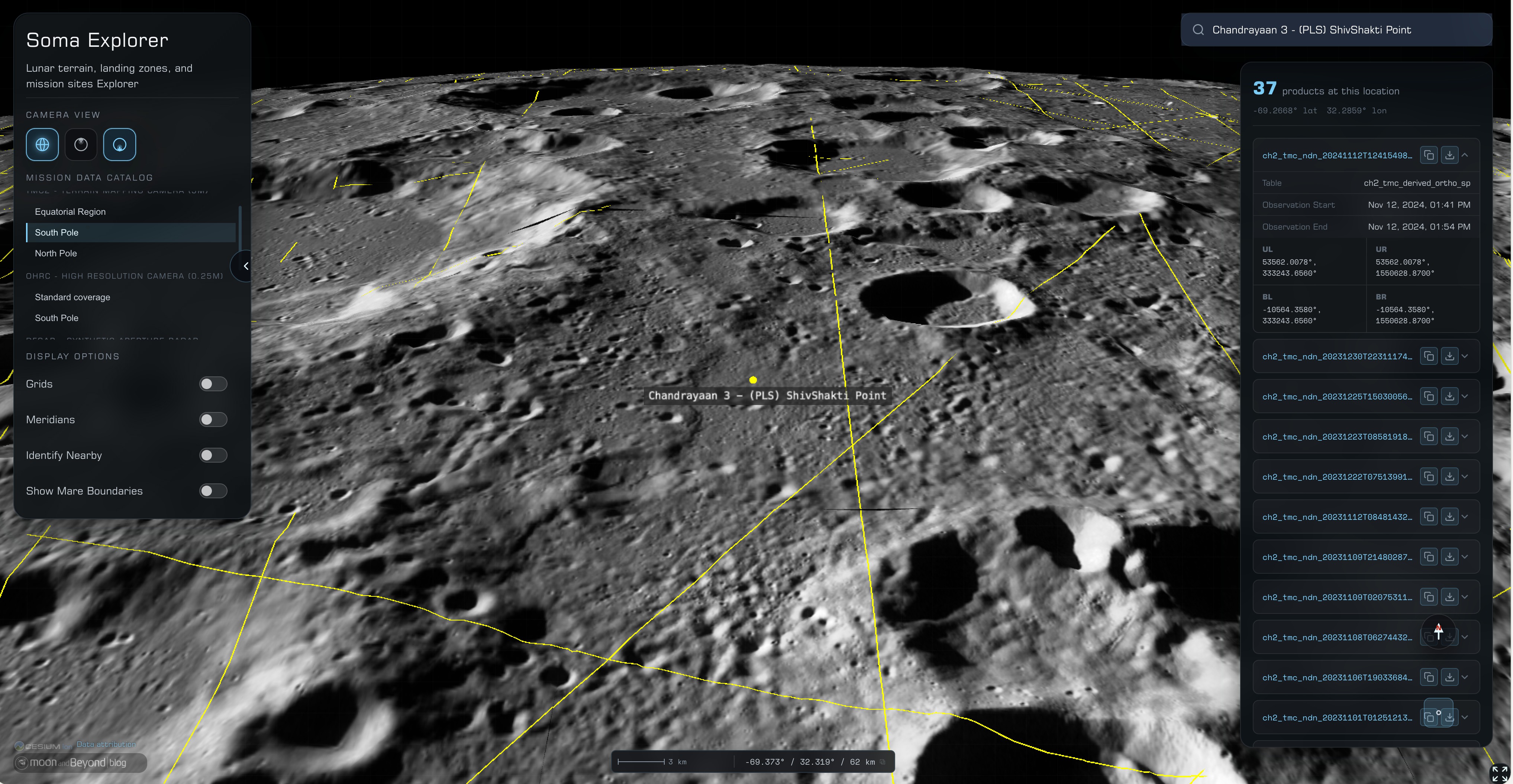}
		\caption{Soma lunar data portal. \textit{Left:} Overview of the OHRC South Pole catalogue showing 203 data products spanning February 2020 to November 2025, with the Chandrayaan-3 landing site (Shiv Shakti Point) marked. \textit{Right:} Location-based query at Shiv Shakti Point returning 37 data products at the selected coordinates, with image footprints displayed on the 3D terrain and individual product metadata (observation times, corner coordinates) available for stereo pair evaluation.}
		\label{fig:soma}
	\end{figure*}

	Preliminary evaluation of Chandrayaan-4 candidate landing sites at Mons Mouton is currently underway using the same stereo pair identification methodology and the DEM generation pipeline described in this paper, informed by the site characterization presented by \citet{Amitabh2026}; results will be reported in a forthcoming publication.

	\subsubsection{Image Selection and Pair Evaluation}

	Using Soma, three OHRC images covering the Shiv Shakti Point region (Chandrayaan-3 Vikram landing site, $\sim$69.37$^\circ$S, 32.35$^\circ$E) were identified, all acquired on April 25, 2024 across three consecutive orbits approximately two hours apart:

	\begin{deluxetable*}{lccc}
		\tablecaption{OHRC Images Used in This Work \label{tab:images}}
		\tablehead{
			\colhead{Parameter} &
			\colhead{Image 1 (Orbit 20812)} &
			\colhead{Image 2 (Orbit 20813)} &
			\colhead{Image 3 (Orbit 20814)}
		}
		\startdata
		Product ID & \texttt{ch2\_ohr\_ncp\_20240425} & \texttt{ch2\_ohr\_ncp\_20240425} & \texttt{ch2\_ohr\_ncp\_20240425} \\
		& \texttt{T1209509264\_d\_img\_d18} & \texttt{T1406019344\_d\_img\_d18} & \texttt{T1603031918\_d\_img\_d18} \\
		Time (UTC) & 12:09:50 & 14:06:01 & 16:03:03 \\
		Altitude & 101.90\,km & 102.35\,km & 101.91\,km \\
		GSD & 0.26\,\mppx & 0.26\,\mppx & 0.26\,\mppx \\
		Pitch & $-10.73^\circ$ & $+9.64^\circ$ & $-12.29^\circ$ \\
		Roll & $-0.07^\circ$ & $+6.63^\circ$ & $+13.29^\circ$ \\
		Dimensions & 12{,}000 $\times$ 93{,}693 & 12{,}000 $\times$ 93{,}693 & 12{,}000 $\times$ 90{,}148 \\
		\hline
		Vikram lander position\tablenotemark{a} & $-$69.5057$^\circ$, 32.3346$^\circ$ & $-$69.5082$^\circ$, 32.3331$^\circ$ & $-$69.5026$^\circ$, 32.2965$^\circ$ \\
		\enddata
		\tablenotetext{a}{SPICE-projected planetocentric coordinates of the Vikram lander pixel in each image, obtained via \texttt{campt}. The spread between images reflects SPICE kernel positioning uncertainty; the $\sim$0.04$^\circ$ longitude offset in Image~3 is consistent with its large roll angle. All positions are $\sim$4\,km from the true landing site, reflecting SPICE kernel uncertainty; the reconstructed DEM exhibits a larger $\sim$6\,km offset after stereo triangulation (Section~\ref{sec:alignment}).}
	\end{deluxetable*}

	All three images were bundle-adjusted together to evaluate stereo pair quality.
	Table~\ref{tab:pairs} shows the match statistics for each pair:

	\begin{deluxetable}{lccc}
		\tablecaption{Stereo Pair Evaluation (3-way Bundle Adjustment) \label{tab:pairs}}
		\tablehead{
			\colhead{Pair} &
			\colhead{Clean Matches} &
			\colhead{Convergence Angle}
		}
		\startdata
		1209 + 1406 & 9{,}918 & 22.66$^\circ$ \\
		1209 + 1603 & 1 & 13.28$^\circ$ \\
		1406 + 1603 & 5 & 24.25$^\circ$ \\
		\enddata
	\end{deluxetable}

	The 1209+1406 pair (orbits 20812/20813) was selected as the only viable stereo pair, with a convergence angle of 22.66$^\circ$ and dense feature correspondence.
	The third image (orbit 20814) had near-zero clean matches with either of the other two, despite adequate convergence angles --- its roll offset ($+13.29^\circ$) placed the region of interest at a different apparent position, preventing overlap in the cropped region.
	A 2$\times$2\,km region of interest (8{,}000 $\times$ 8{,}000 pixels) centered on the Vikram lander was extracted from the selected pair for DEM generation.

	\section{Processing Pipeline}
	\label{sec:pipeline}

	The end-to-end pipeline comprises six stages: ingestion, camera model generation, bundle adjustment, stereo correlation, DEM generation, and post-processing.
	The complete workflow is shown schematically in Figure~\ref{fig:pipeline}.

	The processing environment uses the following software:
	\begin{itemize}
		\item \textbf{ASP} 3.6.0 (build 0fc39ac, December 26, 2025; \citealt{Beyer2018}), pre-built binary tarball from GitHub, bundling Vision Workbench 3.6.0, USGSCSM 2.0.1, and GDAL 3.8.1.
		\item \textbf{ISIS} 2025.11.09 from the \texttt{usgs-astrogeology/label/dev} conda channel (build hc0f069a\_0), running under Rosetta~2 (x86\_64 emulation) on Apple Silicon.
		\item \textbf{ALE} 1.0.2 (commit d7048da), installed from the GitHub \texttt{main} branch.
		The conda-released ALE (0.11.0) lacks the Chandrayaan-2 OHRC driver; the \texttt{main} branch includes it along with geometric fixes from pull request \#682 (sensor frame ID and focal plane mapping corrections).
		\item \textbf{USGSCSM} 3.0.3.3 from conda-forge.
	\end{itemize}
	Camera models were generated using a custom Python script that invokes ALE's OHRC driver to produce ISD files compatible with the CSM framework (\citealt{CSM2015}); the standard \texttt{isd\_generate} tool fails for OHRC because the driver requires SPICE kernels to be pre-loaded into memory, which is not done automatically for Chandrayaan-2 as it is for NASA missions.

	All processing was performed on a single workstation (Apple M3 Max, 16 cores, 128\,GB RAM).
	ISIS and ASP are developed for and natively supported on Linux; the macOS/Rosetta~2 configuration used here introduces no known differences in output quality.

	ASP 3.6.0 includes an experimental Chandrayaan-2 OHRC processing example\footnote{\url{https://stereopipeline.readthedocs.io/en/stable/examples/chandrayaan2.html}} which explicitly states that the workflow ``is not ready for general use and is not reproducible, but is provided for reference.''
	That example requires ISIS and ALE compiled from source, produces a DEM at 1\,m grid spacing with notable jitter artifacts on the order of the image GSD ($\sim$0.25\,m), and requires manual alignment to LOLA after an initial $\sim$4\,km offset was observed.
	The present work addresses these limitations: we achieve 30\,cm grid spacing with 8.1\,cm median triangulation error, using pre-built binaries and conda packages supplemented only by ALE from the \texttt{main} branch and a custom ISD generation script.
	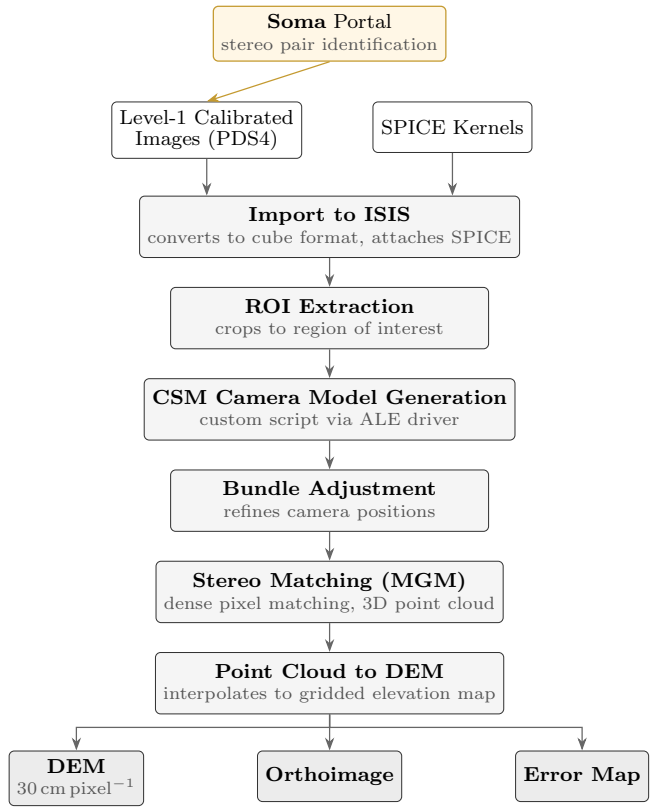
\begin{figure}
		\definecolor{boxgray}{HTML}{333333}
		\definecolor{boxfill}{HTML}{F5F5F5}
		\definecolor{arrowgray}{HTML}{666666}
		\definecolor{descgray}{HTML}{555555}
		\newcommand{\desc}{\scriptsize\color{descgray}}
		\definecolor{somafill}{HTML}{FFF7E6}
		\definecolor{somaborder}{HTML}{C49A30}
		\definecolor{outputfill}{HTML}{ECECEC}
		\centering
		\resizebox{\columnwidth}{!}{%
		\begin{tikzpicture}[
			node distance=0.45cm,
			process/.style={rectangle, draw=boxgray, fill=boxfill, rounded corners=2pt, minimum width=4.4cm, minimum height=0.85cm, align=center, font=\footnotesize, line width=0.4pt},
			input/.style={rectangle, draw=boxgray, fill=white, rounded corners=2pt, minimum width=2.2cm, minimum height=0.7cm, align=center, font=\footnotesize, line width=0.4pt},
			soma/.style={rectangle, draw=somaborder, fill=somafill, rounded corners=2pt, minimum width=2.8cm, minimum height=0.7cm, align=center, font=\footnotesize, line width=0.5pt},
			output/.style={rectangle, draw=boxgray, fill=outputfill, rounded corners=2pt, minimum width=1.8cm, minimum height=0.7cm, align=center, font=\footnotesize, line width=0.4pt},
			myarrow/.style={-{Stealth[length=1.8mm, width=1.3mm]}, arrowgray, line width=0.5pt},
			somaarrow/.style={-{Stealth[length=1.8mm, width=1.3mm]}, somaborder, line width=0.5pt},
			desc/.style={font=\tiny\color{descgray}},
			]
			\node[soma] (soma) {\textbf{Soma} Portal\\[-1pt]{\desc stereo pair identification}};
			\node[input, below=0.55cm of soma, xshift=-1.7cm] (pds4) {Level-1 Calibrated\\[-1pt]Images (PDS4)};
			\node[input, below=0.55cm of soma, xshift=1.7cm] (spice) {SPICE Kernels};
			\draw[somaarrow] (soma.south) -- (pds4.north);
			\node[process, below=0.55cm of $(pds4.south)!0.5!(spice.south)$] (import) {\textbf{Import to ISIS}\\[-1pt]{\desc converts to cube format, attaches SPICE}};
			\draw[myarrow] (pds4.south) -- ([xshift=-1.7cm]import.north);
			\draw[myarrow] (spice.south) -- ([xshift=1.7cm]import.north);
			\node[process, below=0.4cm of import] (crop) {\textbf{ROI Extraction}\\[-1pt]{\desc crops to region of interest}};
			\node[process, below=0.4cm of crop] (camera) {\textbf{CSM Camera Model Generation}\\[-1pt]{\desc custom script via ALE driver}};
			\node[process, below=0.4cm of camera] (bundle) {\textbf{Bundle Adjustment}\\[-1pt]{\desc refines camera positions}};
			\node[process, below=0.4cm of bundle] (stereo) {\textbf{Stereo Matching (MGM)}\\[-1pt]{\desc dense pixel matching, 3D point cloud}};
			\node[process, below=0.4cm of stereo] (dem) {\textbf{Point Cloud to DEM}\\[-1pt]{\desc interpolates to gridded elevation map}};
			\node[output, below left=0.5cm and 0.15cm of dem] (out_dem) {\textbf{DEM}\\[-1pt]{\desc 30\,cm\,pixel$^{-1}$}};
			\node[output, below=0.5cm of dem] (out_ortho) {\textbf{Orthoimage}};
			\node[output, below right=0.5cm and 0.15cm of dem] (out_err) {\textbf{Error Map}};
			\draw[myarrow] (import)--(crop); \draw[myarrow] (crop)--(camera);
			\draw[myarrow] (camera)--(bundle); \draw[myarrow] (bundle)--(stereo);
			\draw[myarrow] (stereo)--(dem);
			\draw[myarrow] (dem.south)--++(0,-0.18)-|(out_dem.north);
			\draw[myarrow] (dem.south)--(out_ortho.north);
			\draw[myarrow] (dem.south)--++(0,-0.18)-|(out_err.north);
		\end{tikzpicture}%
		}
		\caption{Schematic of the OHRC-to-DEM processing pipeline. Stereo pairs are identified using the Soma portal (Section~\ref{sec:data}), and Level-1 calibrated images are ingested into ISIS with SPICE kernels. CSM camera models are generated via a custom ALE integration script, and the Ames Stereo Pipeline performs bundle adjustment, stereo correlation (MGM), and DEM generation. Outputs include a 30\,cm DEM, orthoimage, and triangulation error map.}
		\label{fig:pipeline}
	\end{figure}

	\subsection{Stage 1: Ingestion, SPICE Initialization, and ROI Extraction}

	Level-1 calibrated OHRC images in PDS4 format are converted to ISIS cube format using \texttt{isisimport}.
	Since the images are already radiometrically corrected and geometrically tagged, no additional calibration steps are required at this stage.
	SPICE kernels \citep{Acton1996} --- comprising spacecraft position (SPK), attitude (CK), instrument geometry (IK), and frame definitions (FK) --- are then attached to each image using \texttt{spiceinit} with default settings.
	This requires the correct CK/SPK kernels to already reside in the ISIS data area.
	SPICE data for Chandrayaan-2 are distributed separately on the PRADAN portal and must be manually downloaded and placed in the appropriate directories.
	After adding new kernel files, the ISIS kernel database must be regenerated using \texttt{kerneldbgen}.

	The Vikram lander pixel coordinates were located in each image using \texttt{campt}; the SPICE-projected positions are listed in Table~\ref{tab:images}. The ROI was centered on the Image~1 (orbit 20812) position, and an 8{,}000 $\times$ 8{,}000 pixel region of interest ($\sim$2 $\times$ 2\,km) was extracted from each image using \texttt{crop}.
	Cropping the full strips (12{,}000 $\times$ 93{,}693 pixels) to the region of interest prior to stereo processing reduced total pipeline runtime from hours to approximately 19 minutes.

	\subsection{Stage 2: Camera Model Generation}
	\label{sec:camera}

	Camera model generation required additional integration steps beyond the standard ISIS workflow.
	For most planetary missions supported by ISIS (e.g., LRO, MRO, Cassini), camera model generation is a single automated step using the standard \texttt{isd\_generate} tool.
	The ASP Chandrayaan-2 documentation suggests using \texttt{isd\_generate} for OHRC camera model creation; however, this standard pathway does not reliably produce working camera models for OHRC data in practice.

	We addressed this with a custom Python script that generates CSM-compatible ISD (Instrument Support Data) files.
	The script reads kernel paths from the ISIS cube label, loads all SPICE kernels into memory, invokes ALE's OHRC camera driver directly, and writes the resulting camera model to disk.
	This manual kernel loading is necessary because ALE does not automatically locate Chandrayaan-2 kernels as it does for NASA missions, where kernel paths are pre-configured in the ISIS data area.

	The choice of CSM over the legacy ISIS camera model proved to be important, as discussed in Section~\ref{sec:csm_comparison}.

	\subsection{Stage 3: Bundle Adjustment}
	\label{sec:bundle}

	Bundle adjustment refines camera positions and orientations by minimizing reprojection error across matched features in the stereo pair.
	This was performed using ASP's \texttt{bundle\_adjust} on all three images simultaneously (3-way adjustment), with 500 interest points per tile as the only non-default parameter; all other settings used ASP defaults (DENSE\_SCHUR solver, Levenberg--Marquardt trust region, automatic outlier removal, 2 passes).
	The adjustment converged in two passes: Pass~1 ran 43 iterations reducing the cost function from 25{,}512 to 1{,}115 (95.6\% reduction); Pass~2 converged in 2 iterations, confirming stability.
	A total of 279 outliers (1.4\%) were removed during the process.

	\subsubsection{CSM vs.\ ISIS Camera Model Comparison}
	\label{sec:csm_comparison}

	The importance of CSM camera models was established during a proof-of-concept evaluation on a separate OHRC stereo pair covering the Chandrayaan-3 candidate site designated S7 by \citet{Amitabh2021}, using the same stereo pair listed in their Table~2.
	This site was chosen as an initial test case to validate the end-to-end pipeline before applying it to the primary target at Shiv Shakti Point.
	We evaluated both the legacy ISIS line-by-line camera model and the CSM continuous interpolation model on that pair.
	Both camera model configurations were processed under identical stereo parameters, interest point detection settings, and bundle adjustment configurations within the Ames Stereo Pipeline to ensure a controlled comparison.
	The results are summarized in Table~\ref{tab:csm_comparison}.

	\begin{deluxetable}{lcc}
		\tablecaption{Camera Model Comparison (S7 Site Evaluation) \label{tab:csm_comparison}}
		\tablehead{
			\colhead{Metric} &
			\colhead{ISIS Camera} &
			\colhead{CSM Camera}
		}
		\startdata
		Feature matches & $\sim$122 & $\sim$94{,}000 \\
		IP coverage & Sparse & $>$97\% \\
		Disparity search range & 9{,}500 $\times$ 370\,px & 335 $\times$ 93\,px \\
		Processing time per tile & 15--30\,min & $\sim$20\,sec \\
		Total stereo time & $>$10\,hr & $\sim$30\,min \\
		\enddata
	\end{deluxetable}

	The CSM-based path yielded approximately three orders of magnitude more feature matches, which in turn substantially constrained the disparity search range and reduced per-tile stereo processing time by a factor of 20--50$\times$.
	The ISIS camera path produced so few matches that the resulting DEM was unreliable.
	The primary advantage of the CSM path was stable convergence and successful DEM reconstruction; reduced runtime is a secondary consequence of improved match density and solver stability.
	The improved performance of the CSM implementation likely reflects more accurate time-dependent sensor modeling and line-scanning geometry handling relative to the legacy ISIS camera model, which uses a line-by-line discrete approximation that may not adequately capture the continuous attitude variations introduced by OHRC's pitch-maneuver stereo acquisition.

	In our experiments across two OHRC stereo pairs, the legacy ISIS camera model failed to achieve stable convergence, whereas CSM-based modeling produced robust solutions.
	This finding was established on the S7 proof-of-concept and confirmed on the Shiv Shakti Point stereo pair presented in this paper.
	The sample size is limited to two stereo pairs, and the result may reflect the specific combination of OHRC pushbroom geometry and current SPICE kernel quality rather than a universal limitation of the ISIS camera model.
	All experiments were conducted using the toolchain versions listed in Section~\ref{sec:pipeline}; results may evolve with future sensor model updates.
	An open pull request on the ALE repository\footnote{\url{https://github.com/DOI-USGS/ale/pull/682}} includes geometric corrections to the OHRC driver that should further improve CSM-based results.

	\subsection{Stage 4: Stereo Correlation}

	Dense stereo matching was performed using ASP's \texttt{parallel\_stereo} with the MGM (More Global Matching) algorithm \citep{Hirschmuller2005}, which is well-suited to pushbroom line-scan geometry.
	Bundle-adjusted match files were reused from Stage~3, and a conservative per-tile correlation timeout was applied.
	Processing is parallelized on a per-tile basis, with results blended to produce a continuous disparity map.
	Full implementation details and workarounds for ASP integration issues are documented in the supplementary notes \citep{\suppref\suppyear}.

	\subsection{Stage 5: DEM Generation}

	The triangulated 3D point cloud was interpolated into a regular gridded DEM using ASP's \texttt{point2dem}, generating a per-pixel triangulation error map and terrain-corrected orthoimage alongside the elevation raster.
	Grid spacing was determined automatically from point cloud density, yielding $\sim$0.303\,\mppx for this dataset.
	Forcing finer spacing would produce interpolated artifacts without additional topographic information.

	Three products were generated:
	\begin{itemize}
		\item Digital Elevation Model (elevation relative to the lunar reference ellipsoid, $R$ = 1737.4\,km)
		\item Orthorectified image (terrain-corrected, map-projected)
		\item Triangulation error map (per-pixel intersection error)
	\end{itemize}

	\subsection{Stage 6: Post-processing and Visualization}

	Lunar elevation values are referenced to the 1737.4\,km sphere and therefore appear as large negative numbers (typically $-$1500 to $-$1800\,m).
	For visualization and compatibility with standard GIS and 3D rendering tools, we applied a datum shift to produce positive elevation values. Final products were rendered using QGIS (Figures~\ref{fig:3d_multi} and~\ref{fig:3d_oblique}).

	\subsection{Slope Computation}

	The DEM was generated using ASP's \texttt{point2dem}, which automatically selected an oblique stereographic projection centered on the median latitude and longitude of the reconstructed point cloud (origin: 69.177$^\circ$S, 32.326$^\circ$E).
	The projection assumes a spherical Moon with radius $R$ = 1737.4\,km (eccentricity = 0) and uses linear units of meters.
	The DEM grid spacing is 0.303\,\mppx.

	Slope was computed in degrees using GDAL's \texttt{gdaldem slope} implementation (Horn 3$\times$3 finite-difference kernel, \texttt{-{}-compute\_edges}) with a $Z$-factor of 1, as both horizontal and vertical units are meters.
	Percentiles and area fractions were computed over valid pixels only.
	No additional smoothing or filtering was applied prior to slope calculation.
	Over the $\sim$2\,km footprint, projection-induced scale distortion is negligible relative to pixel resolution.

	\subsection{Processing Time Summary}

	Table~\ref{tab:timing} summarizes the wall-clock time for each pipeline stage on the 2 $\times$ 2\,km ROI crop, measured on the Apple M3 Max workstation.

	\begin{deluxetable}{lc}
		\tablecaption{Pipeline Timing (Vikram ROI, 8{,}000 $\times$ 8{,}000 px crop) \label{tab:timing}}
		\tablehead{\colhead{Stage} & \colhead{Duration}}
		\startdata
		Bundle adjustment (3 images) & $\sim$3.7\,min \\
		Preprocessing & 13\,sec \\
		Correlation + blend + subpixel & $\sim$7\,min \\
		Filtering & 16\,sec \\
		Triangulation & $\sim$3.7\,min \\
		DEM generation (\texttt{point2dem}) & $\sim$1.4\,min \\
		\hline
		Total & $\sim$19\,min \\
		\enddata
	\end{deluxetable}

	Cropping to the region of interest prior to stereo processing was essential; full-strip processing (12{,}000 $\times$ $\sim$93{,}000 pixels) requires 16--17 hours on the same hardware.

	\section{Results}
	\label{sec:results}

	The resulting DEM covers the Chandrayaan-3 Vikram landing site with the parameters summarized in Table~\ref{tab:dem_results}.

	\begin{deluxetable}{lc}
		\tablecaption{DEM Product Summary \label{tab:dem_results}}
		\tablehead{\colhead{Metric} & \colhead{Value}}
		\startdata
		Horizontal resolution & 0.303\,\mppx ($\sim$30\,cm) \\
		DEM dimensions & 7{,}193 $\times$ 7{,}381 pixels \\
		Ground coverage & 2.18 $\times$ 2.24\,km \\
		Valid pixel fraction & 91.2\% \\
		Elevation range & $-$277.6\,m to $-$220.8\,m \\
		Total relief & 56.9\,m \\
		Median triangulation error & 0.081\,m (8.1\,cm) \\
		P95 triangulation error & 0.256\,m (25.6\,cm) \\
		P99 triangulation error & 0.339\,m (33.9\,cm) \\
		\enddata
	\end{deluxetable}

	For context, this resolution is approximately 3$\times$ finer than LOLA's best single-track measurements ($\sim$1\,m at select sites) and over 100$\times$ finer than LOLA's global gridded product ($\sim$30\,m).
	It is comparable to or better than the resolution reported by SAC using the proprietary OPTIMUS pipeline (28\,cm horizontal, 100\,cm vertical; \citealt{Amitabh2021}), with our estimated relative vertical precision (40--50\,cm) comparable in magnitude, noting differences in methodology and the absence of direct cross-validation.
	This precision estimate is derived from the triangulation error distribution and stereo convergence geometry, and reflects relative precision at sub-meter scale rather than absolute accuracy with respect to an external datum.

	At this resolution, individual boulders, meter-scale craters, and subtle regolith texture variations are clearly resolved in the elevation data.
	The Chandrayaan-3 Vikram lander itself appears as a distinct elevation feature --- a bright bump on the surface --- resolved in both the orthoimage and the DEM (Figure~\ref{fig:roi}).
	The Pragyan rover is also visible approximately 8.3\,m to the upper left of the lander, with both objects casting shadows consistent in direction and length with the solar incidence angle at the time of acquisition.
	The terrain surrounding the landing site is confirmed to be remarkably flat, consistent with the site selection criteria for Chandrayaan-3: the color-coded elevation map (Figure~\ref{fig:dem}) shows widely spaced contours in the immediate vicinity of the lander, with elevation varying by less than 3\,m across several hundred meters.

	\subsection{Terrain Statistics}

	Slope statistics derived from the 0.30\,\mppx DEM are summarized in Table~\ref{tab:slopes} for both the full DEM footprint (F1; 2.18 $\times$ 2.24\,km) and a 250\,m radius buffer centered on the Vikram lander (F2).
	Over the full footprint, the mean slope is 3.61$^\circ$ ($\sigma$ = 2.93$^\circ$), with a 95th percentile slope of 9.25$^\circ$.
	Within the 250\,m lander buffer, the mean slope decreases to 3.10$^\circ$ and the 95th percentile slope to 7.24$^\circ$, indicating that the immediate landing zone is statistically flatter than the broader site.
	Only 1.5\% of terrain within the lander buffer exceeds 10$^\circ$, and 0.2\% exceeds 15$^\circ$.

	These values are comparable in magnitude to the $\sim$4.27$^\circ$ global mean slope reported for LS-2 by \citet{Amitabh2023}, noting differences in DEM generation and slope computation methodology.
	The derived slope field therefore exhibits terrain characteristics consistent with the previously reported low-relief nature of the Chandrayaan-3 landing region.

	\begin{deluxetable*}{lcc}
		\tablecaption{Terrain and Slope Statistics \label{tab:slopes}}
		\tablehead{
			\colhead{Metric} &
			\colhead{F1 (2.18 $\times$ 2.24\,km)} &
			\colhead{F2 (250\,m buffer)}
		}
		\startdata
		Relief (m) & 56.9 & 13.7 \\
		Mean slope ($^\circ$) & 3.61 & 3.10 \\
		Median slope ($^\circ$) & 2.83 & 2.62 \\
		Std dev slope ($^\circ$) & 2.93 & 2.17 \\
		P90 slope ($^\circ$) & 7.20 & 5.67 \\
		P95 slope ($^\circ$) & 9.25 & 7.24 \\
		P99 slope ($^\circ$) & 13.98 & 10.98 \\
		\% area $>$ 5$^\circ$ & 20.9\% & 13.8\% \\
		\% area $>$ 10$^\circ$ & 3.9\% & 1.5\% \\
		\% area $>$ 15$^\circ$ & 0.7\% & 0.2\% \\
		\enddata
	\end{deluxetable*}

	\subsection{Elevation and Reconstruction Quality}

	Elevation and triangulation error statistics are summarized in Table~\ref{tab:elevation}.
	The full-footprint relief is 56.9\,m, with a standard deviation of 8.37\,m.
	Within the 250\,m lander buffer, local relief is 13.7\,m and elevation standard deviation decreases to 2.62\,m, consistent with the visually observed low-relief terrain near the landing site.

	The median triangulation error across both regions is 8.1\,cm.
	The 95th percentile triangulation error decreases from 25.6\,cm over the full footprint to 16.7\,cm within the lander buffer, indicating that the landing site lies within one of the lowest-error regions of the reconstruction.
	These statistics support the internal geometric stability of the stereo solution.

	\begin{deluxetable*}{lcc}
		\tablecaption{Elevation Distribution and Triangulation Error Statistics \label{tab:elevation}}
		\tablehead{
			\colhead{Metric} &
			\colhead{F1 (2.18 $\times$ 2.24\,km)} &
			\colhead{F2 (250\,m buffer)}
		}
		\startdata
		Min elevation (m) & $-$277.6 & $-$251.5 \\
		Max elevation (m) & $-$220.8 & $-$237.8 \\
		Mean elevation (m) & $-$243.2 & $-$244.7 \\
		Median elevation (m) & $-$243.6 & $-$244.8 \\
		Std dev elevation (m) & 8.37 & 2.62 \\
		Valid pixels (\%) & 91.2 & 100.0 \\
		Median triangulation error (cm) & 8.1 & 8.1 \\
		P95 triangulation error (cm) & 25.6 & 16.7 \\
		P99 triangulation error (cm) & 33.9 & 19.8 \\
		\enddata
	\end{deluxetable*}

	\subsection{Absolute Geolocation Considerations}

	The unaligned stereo-derived DEM exhibits a systematic geolocation offset of $\sim$5.97\,km relative to published Chandrayaan-3 landing coordinates and to single-image SPICE-projected positions obtained via \texttt{campt}.
	Similar kilometer-scale offsets have been reported in the experimental OHRC example provided in the ASP documentation, where the DEM required manual alignment to LOLA after an initial $\sim$4\,km shift.

	These offsets arise because bundle adjustment was performed without external ground control points and therefore refines camera geometry in a relative sense only.
	Pushbroom stereo acquired with along-track pitch maneuvers is particularly sensitive to small spacecraft attitude and position errors in the released SPICE kernels.
	The analysis was performed using publicly released SPICE CK and SPK kernels without reconstructed or refined state vector adjustments.
	Geodetic alignment is described in Section~\ref{sec:alignment}.

	\begin{figure*}
		\centering
		\includegraphics[width=0.48\textwidth]{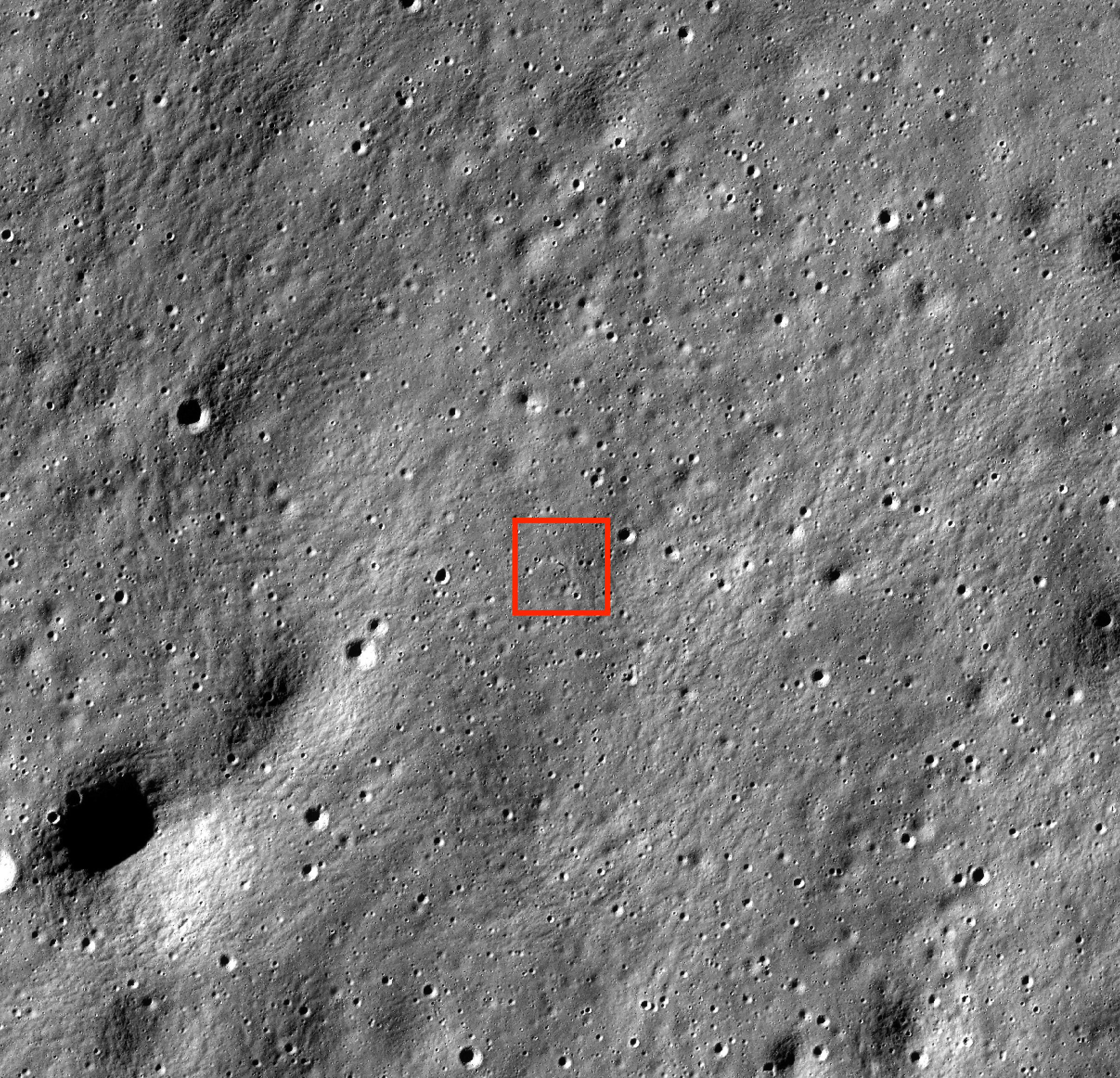}\hfill
		\includegraphics[width=0.48\textwidth]{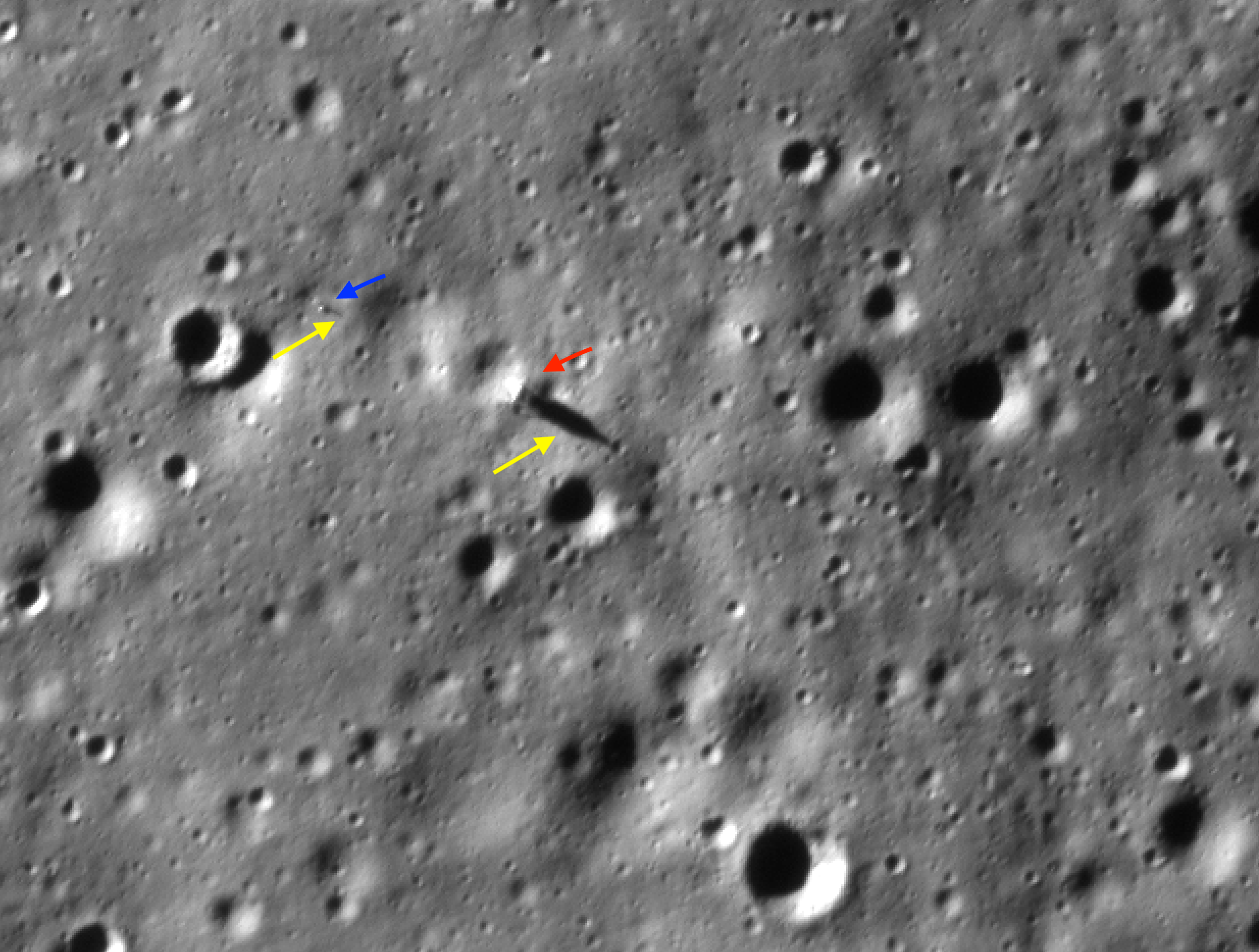}
		\caption{OHRC orthoimage of the Shiv Shakti Point region from image \texttt{ch2\_ohr\_ncp\_20240425T1406019344} (orbit 20813). \textit{Left:} The full 2 $\times$ 2\,km region of interest used for DEM generation. The red box marks the SPICE-projected location of the Chandrayaan-3 Vikram lander (69.5082$^\circ$S, 32.3331$^\circ$E; Table~\ref{tab:images}), prior to geodetic alignment. \textit{Right:} Zoomed view showing the Vikram lander (red arrow) and the Pragyan rover (blue arrow) approximately 8.3\,m to the upper left (69.5079$^\circ$S, 32.3331$^\circ$E), consistent with the rover traverse documented by \citet{Iyer2025}. Yellow arrows indicate the shadows cast by both objects, which are consistent in direction and length, confirming the solar incidence angle at the time of acquisition ($\sim$78$^\circ$). Both the lander and rover are resolved as distinct features at the 0.26\,\mppx GSD. Image credit: \authorcredit.}
		\label{fig:roi}
	\end{figure*}

	\begin{figure*}
		\centering
		\includegraphics[width=\textwidth]{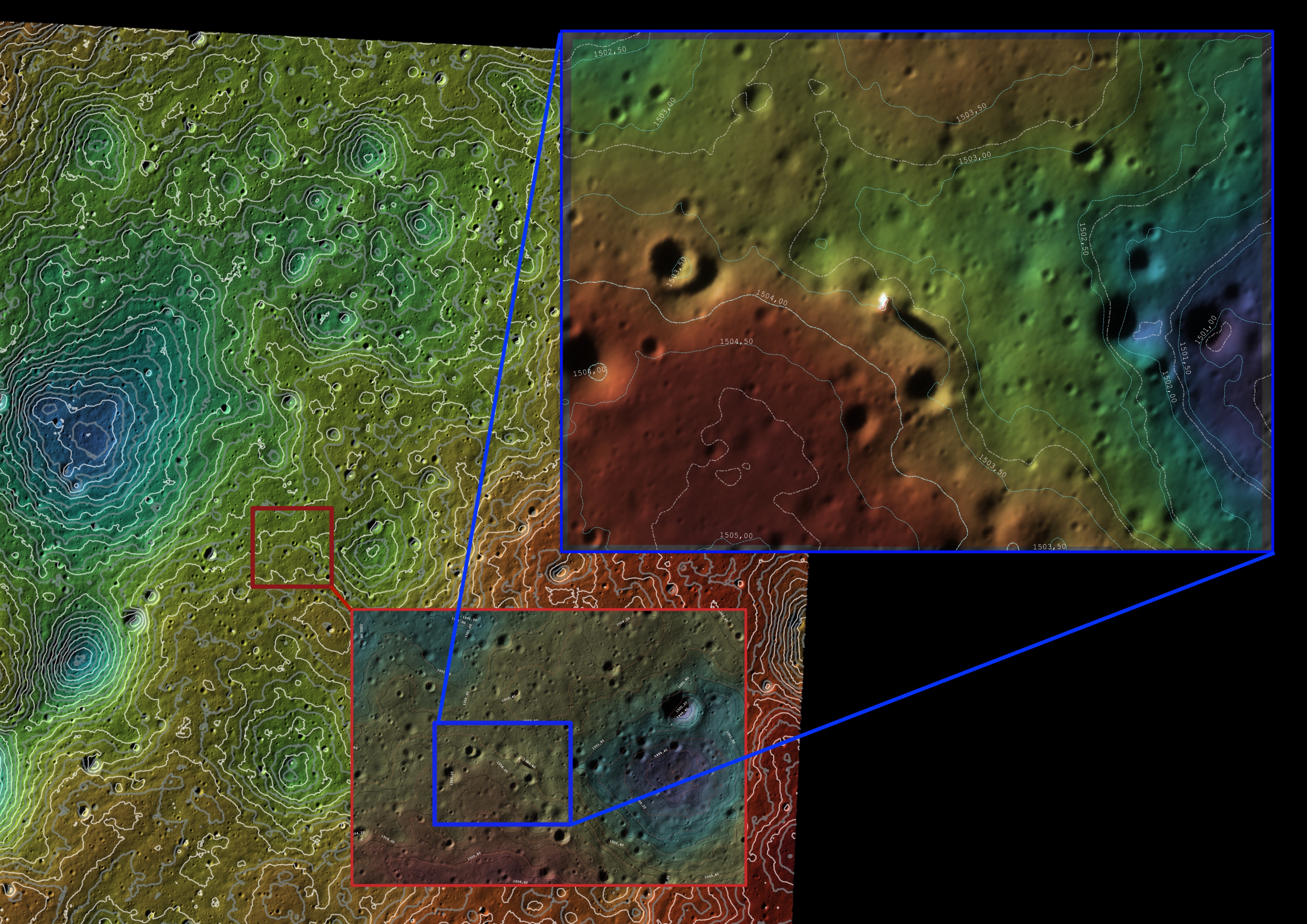}
		\caption{Color-coded elevation map of the Shiv Shakti Point DEM with 50\,cm contour intervals, shown at three zoom levels. \textit{Left:} Full 2.18 $\times$ 2.24\,km DEM extent; the color scale spans approximately 57\,m of total relief. \textit{Red inset:} Intermediate zoom of the Vikram landing area, showing widely spaced contours that confirm the exceptional flatness of the selected landing site --- elevation varies by less than 3\,m across several hundred meters. \textit{Blue inset:} Fine-scale zoom showing sub-meter contour detail around individual craters and the lander location. Elevations are datum-shifted to positive values (referenced to the 1737.4\,km lunar sphere). DEM shown in oblique stereographic projection centered on the scene. Image credit: \authorcredit.}
		\label{fig:dem}
	\end{figure*}

	\begin{figure*}
		\centering
		\includegraphics[width=\textwidth]{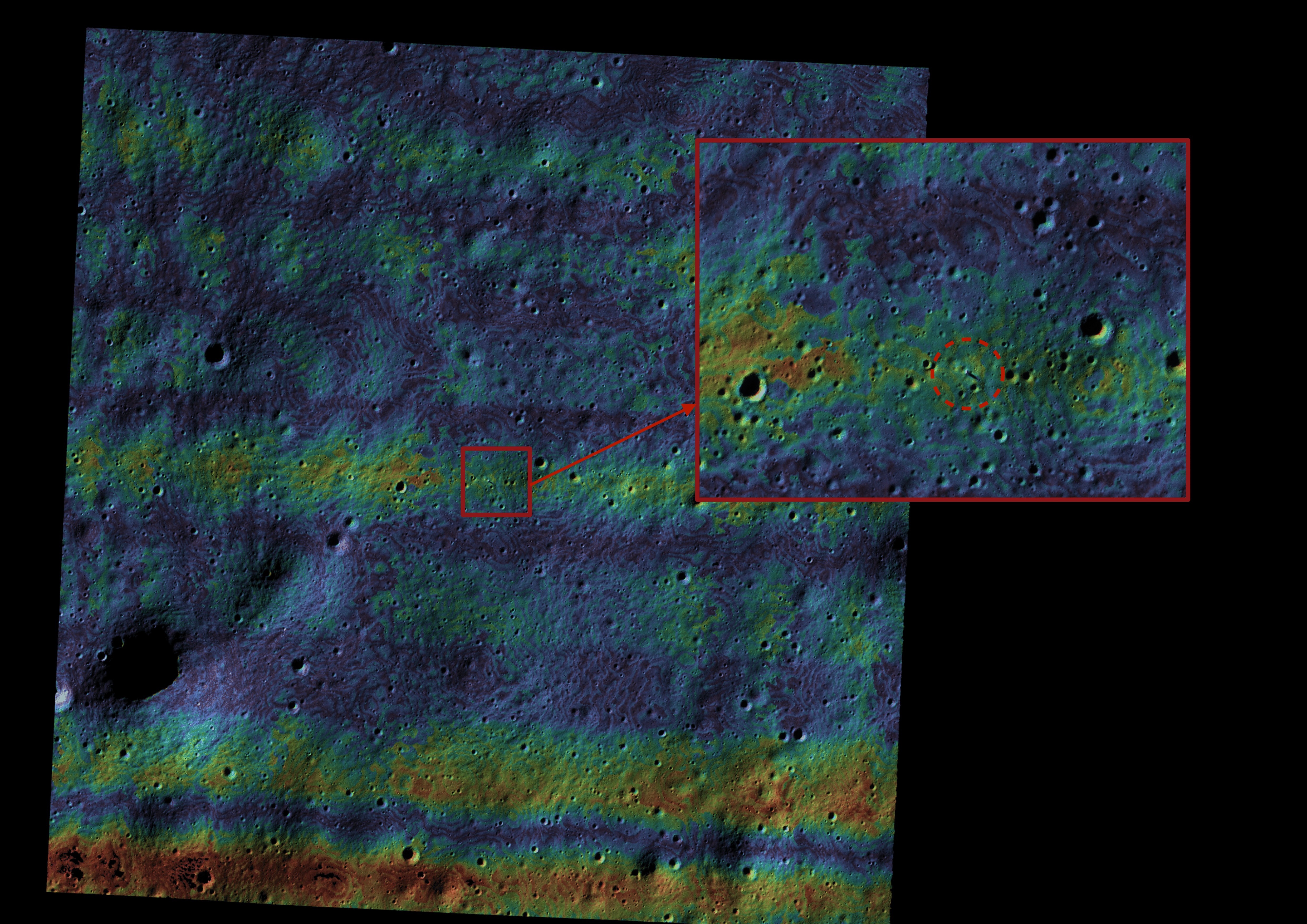}
		\caption{Triangulation error map overlaid on the orthoimage. Blue/dark tones indicate low intersection error ($<$5\,cm); green/yellow indicate moderate error (5--15\,cm); warm tones (orange/red) indicate elevated error ($>$20\,cm). The majority of the scene is dominated by low error (median 8.1\,cm), with elevated values concentrated along crater walls and rims (where stereo occlusion and shadow reduce matching quality), at strip margins (weaker stereo geometry), and in along-track bands (a known ASP MGM tile-blending artifact). \textit{Red inset:} Zoomed view of the Vikram lander area (dashed circle), which sits in a low-error zone consistent with the flat, well-illuminated terrain at the landing site. Image credit: \authorcredit.}
		\label{fig:error}
	\end{figure*}

	\begin{figure*}
		\centering
		\includegraphics[width=\textwidth]{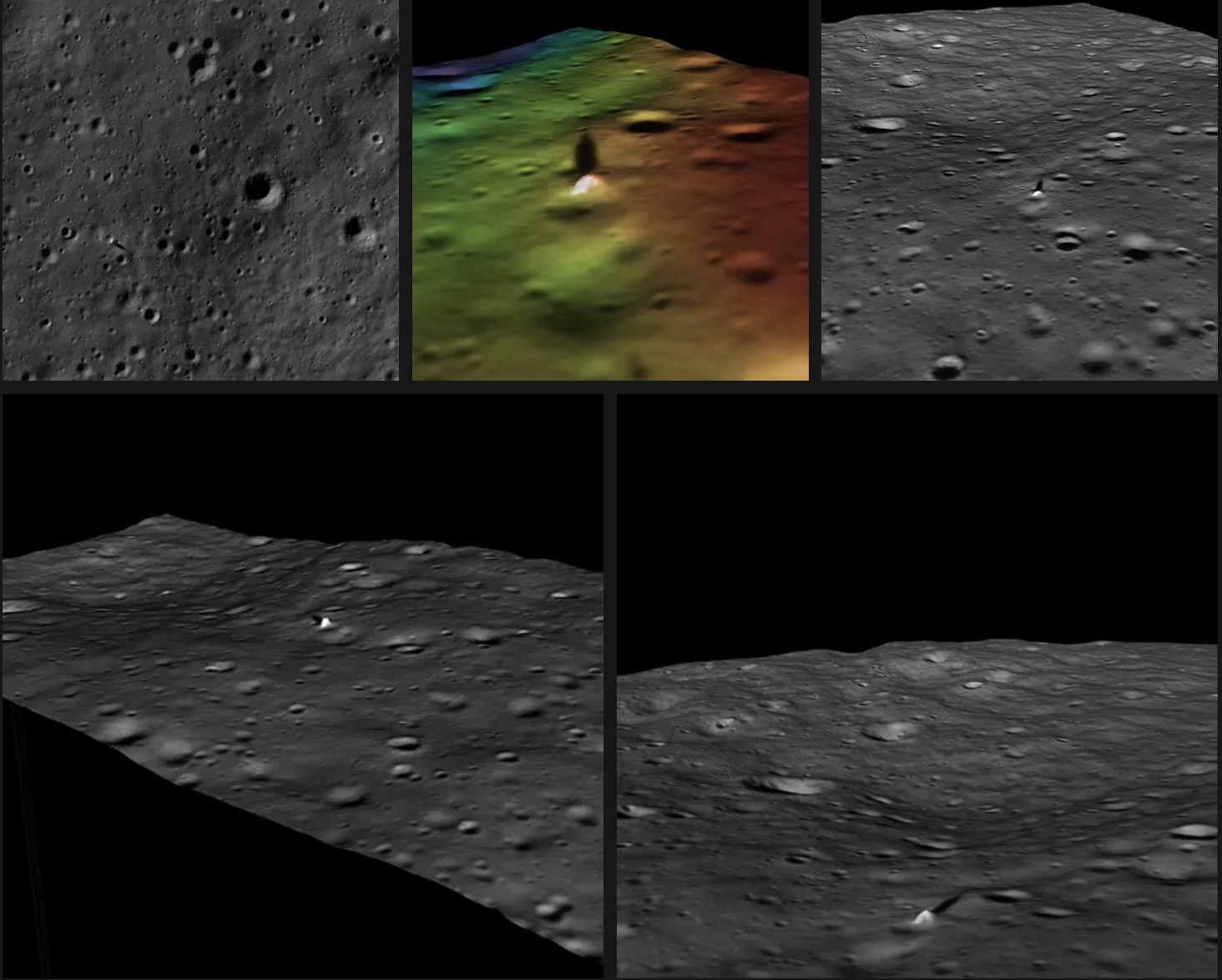}
		\caption{Three-dimensional perspective views of the Shiv Shakti Point DEM rendered in QGIS using the Qgis2threejs plugin \citep{qgis2threejs}. \textit{Top left:} Nadir orthoimage. \textit{Top center:} Oblique view with color-coded elevation draped on the 3D surface; the Vikram lander is visible as the bright feature at center. \textit{Top right:} Oblique view with the orthoimage draped on the DEM. \textit{Bottom:} Two additional perspective angles with orthoimage drape, showing the terrain context and the gentle topographic gradients surrounding the landing site. Vertical exaggeration has been applied to emphasize surface relief. Image credit: \authorcredit.}
		\label{fig:3d_multi}
	\end{figure*}

	\begin{figure*}
		\centering
		\includegraphics[width=\textwidth]{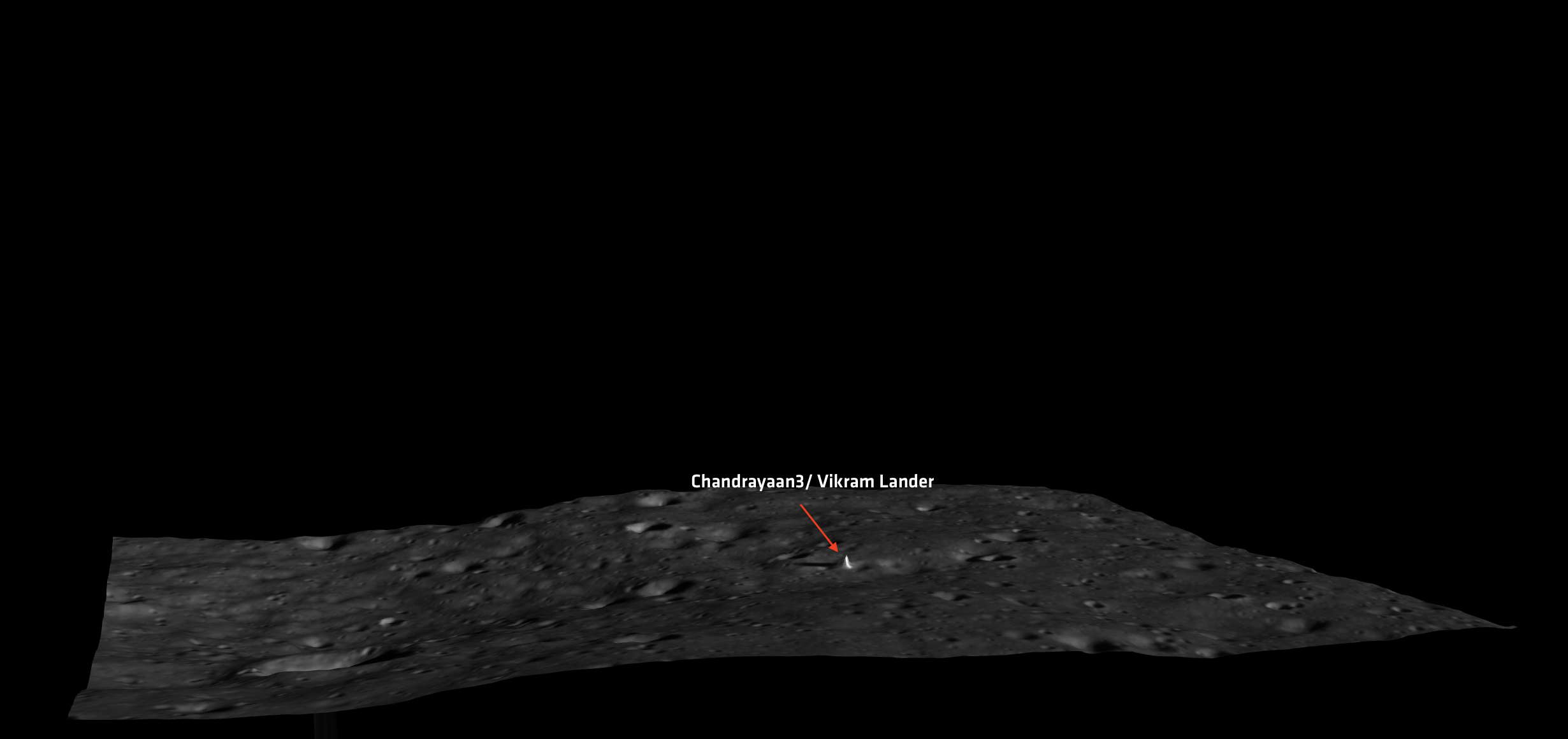}
		\caption{Wide oblique 3D perspective of the DEM with orthoimage drape, showing the full extent of the reconstructed terrain. The Chandrayaan-3 Vikram lander (red arrow) is resolved as a distinct bright feature on the surface. Vertical exaggeration has been applied to emphasize terrain features; the actual surface is substantially flatter than depicted, with only $\sim$57\,m of total relief across the 2.18 $\times$ 2.24\,km coverage area. Without exaggeration, craters, ridges, and subtle topographic variations would be nearly imperceptible at this viewing scale. Rendered using the Qgis2threejs plugin. Image credit: \authorcredit.}
		\label{fig:3d_oblique}
	\end{figure*}

	\section{Geodetic Alignment}
	\label{sec:alignment}

	\subsection{Reference Data --- LOLA}

	Our initial approach used Lunar Orbiter Laser Altimeter (LOLA) Reduced Data Record (RDR) point measurements obtained via the NASA Orbital Data Explorer (ODE; \url{https://ode.rsl.wustl.edu}), following the robust altimetric co-registration methodology established for LRO data products \citep{Barker2016,Mazarico2012}.
	Table~\ref{tab:lola_selection} summarizes the selection parameters.

	\begin{deluxetable}{lc}
		\tablecaption{LOLA RDR Data Selection Parameters \label{tab:lola_selection}}
		\tablehead{\colhead{Parameter} & \colhead{Value}}
		\startdata
		Latitude range & $-$69.6$^\circ$S to $-$69.0$^\circ$S \\
		Longitude range & 32$^\circ$E--33$^\circ$E (0--360$^\circ$ system) \\
		Coordinate system & Planetocentric \\
		Datum & D\_MOON ($R$ = 1{,}737{,}400\,m) \\
		Total LOLA shots intersecting DEM & 34{,}190 \\
		LOLA product used & RDR TopoFull CSV \\
		\enddata
	\end{deluxetable}

	\subsection{Initial Approach: LOLA Alignment (Failed)}
	\label{sec:lola_failed}

	Direct alignment of the 0.30\,m DEM to raw LOLA shot data using ASP's \texttt{pc\_align} produced large residuals (median $\sim$11\,km) and did not converge reliably, owing to the resolution mismatch between the DEM and the sparse LOLA sampling.
	We therefore adopted a multi-resolution approach: LOLA shots were first clipped to the DEM spatial extent, then gridded to a common 20\,m resolution using ASP's \texttt{point2dem} with a search radius factor of 4.
	The OHRC DEM was downsampled to 20\,m using \texttt{gdalwarp} with average resampling.
	Table~\ref{tab:gridding} summarizes the gridding specifications.

	A grid spacing of 20\,m was selected to avoid over-interpolation between sparse LOLA tracks while preserving true altimetric structure; 5\,m spacing was tested but rejected as it introduced interpolation artifacts.

	\begin{deluxetable*}{lccc}
		\tablecaption{DEM and LOLA Gridding Specifications \label{tab:gridding}}
		\tablehead{
			\colhead{Product} &
			\colhead{Resolution} &
			\colhead{Projection} &
			\colhead{Valid Pixel Coverage}
		}
		\startdata
		OHRC DEM (original) & 0.303\,m & Oblique stereographic & 91.58\% \\
		OHRC DEM (downsampled) & 20\,m & Oblique stereographic & --- \\
		LOLA gridded DEM & 20\,m & Oblique stereographic & 83.66\% \\
		\enddata
		\tablecomments{Projection parameters: stereographic, latitude of origin $-$69.1772$^\circ$, longitude of origin 32.3255$^\circ$, $R$ = 1{,}737{,}400\,m, false easting/northing = 0\,m.}
	\end{deluxetable*}

	Translation-only \texttt{pc\_align} applied a 1{,}164\,m shift.
	A subsequent rigid ICP refinement (point-to-plane) introduced a 0.81$^\circ$ rotation.
	However, post-alignment verification revealed that the LOLA alignment \emph{failed} to constrain horizontal position.
	The Vikram lander feature, which could be located in the DEM by visual inspection, moved from 5.97\,km to 7.78\,km from ISRO's announced coordinates \citep{ISRO2023} after alignment --- the procedure made horizontal positioning \emph{worse}.

	The failure is attributable to the nature of the terrain and the reference data.
	At this site, the terrain within the OHRC footprint is exceptionally gentle and featureless at the 20\,m LOLA grid scale: widely spaced contours with no sharp ridges, escarpments, or distinctive crater morphology that would constrain horizontal registration.
	The \texttt{pc\_align} ICP algorithm minimizes elevation differences (point-to-surface distances), which effectively constrains the vertical ($Z$) component of the transform but provides little leverage on horizontal ($XY$) position when the reference surface lacks lateral topographic gradients.
	The resulting translation was $Z$-optimal but $XY$-unconstrained, and the rigid rotation overfitted noise in the smooth LOLA surface, further degrading the horizontal solution.

	This result demonstrates that LOLA-based alignment is insufficient for sites with gentle, featureless terrain at decameter scales, a finding that may be relevant to other high-latitude landing site DEMs where similar terrain conditions prevail.

	\subsection{Reference Data --- LROC NAC DEM}

	Given the failure of LOLA-based alignment, we adopted an LROC Narrow Angle Camera (NAC) stereo DEM as the alignment reference.
	The product \texttt{NAC\_DTM\_VIKRAMSITE1}\footnote{\url{https://data.lroc.im-ldi.com/lroc/view_rdr/NAC_DTM_VIKRAMSITE1}} covers the Chandrayaan-3 landing site at approximately 3\,m\,pixel$^{-1}$ resolution, derived from NAC stereo pairs using the methodology described by \citet{Henriksen2016} and distributed through the PDS Geosciences Node.
	The product has a relative linear error (LE) of 1.92\,m and was registered to 63 LOLA profiles with an RMS of 1.85\,m.
	At 3\,m resolution, the LROC NAC DEM resolves craters, ridges, and terrain features that provide strong constraints on all three spatial axes, unlike the smooth 20\,m LOLA surface.

	The OHRC DEM uses an oblique stereographic projection centered on the scene (latitude of origin $-$69.1772$^\circ$, longitude of origin 32.3255$^\circ$), whereas the LROC NAC DEM uses a polar stereographic projection (latitude of origin $-$90$^\circ$).
	This CRS difference required careful reprojection for pixel-wise comparison (Section~\ref{sec:reproj}).

	\subsection{LROC NAC Alignment}
	\label{sec:nac_alignment}

	\subsubsection{Step 1: Automated Point-Cloud Registration}

	An initial automated alignment was performed using \texttt{pc\_align} in translation-only mode between the OHRC DEM and a 3\,km crop of the LROC NAC DEM centered on the landing site.
	This produced a 5{,}294\,m translation, reducing the offset from the Vikram lander to approximately 750\,m.
	While a significant improvement over the raw 5.97\,km offset, the residual was still too large for geodetic-quality registration.

	\subsubsection{Step 2: Manual Tie-Point Refinement}

	To achieve sub-hundred-meter accuracy, five crater centers were identified as tie-points in both the OHRC and LROC NAC DEMs using the ISIS \texttt{qview} tool.
	From these correspondences, the mean geographic offset was computed as $\Delta\mathrm{lat} \approx -0.197^\circ$, $\Delta\mathrm{lon} \approx -0.007^\circ$, with a root-mean-square deviation (RMSD) of tie-point residuals of 12.3\,m.
	The corresponding ECEF translation was encoded as a $4 \times 4$ affine matrix and applied to the full-resolution OHRC point cloud via \texttt{pc\_align} with \texttt{-{}-num-iterations~0} (transform propagation only, no further optimization).

	After tie-point alignment, the Vikram lander feature in the OHRC DEM is located $\sim$30\,m from ISRO's announced landing coordinates ($-$69.3733$^\circ$S, 32.3186$^\circ$E).
	The per-point vertical offsets between the tie-point pairs show a mean $\Delta z$ of $+$780\,m, reflecting the absolute vertical positioning error inherited from the SPICE kernels.
	The 3D alignment transform corrects both the $\sim$5.97\,km horizontal offset and the $\sim$780\,m vertical offset simultaneously.
	Table~\ref{tab:alignment_params} summarizes the final alignment parameters.

	\begin{deluxetable}{lc}
		\tablecaption{Final Alignment Parameters (LROC NAC Tie-Point) \label{tab:alignment_params}}
		\tablehead{\colhead{Parameter} & \colhead{Value}}
		\startdata
		Translation magnitude (3D) & 6{,}021\,m \\
		Translation (NED) & N: $-$5{,}971\,m; E: $-$71\,m; D: $-$769\,m \\
		Number of tie-points & 5 (crater centers) \\
		Tie-point RMSD & 12.3\,m \\
		Final offset from ISRO coordinates & $\sim$30\,m \\
		Reference dataset & LROC NAC DEM ($\sim$3\,m\,pixel$^{-1}$) \\
		\enddata
	\end{deluxetable}

	\subsection{Alignment Summary}
	\label{sec:alignment_summary}

	Table~\ref{tab:alignment_attempts} presents the complete alignment history, showing the measured Vikram lander position and offset from ISRO's announced coordinates at each stage.
	The LOLA-based approach optimized vertical fit but degraded horizontal position, while the LROC NAC alignment progressively reduced the offset from 5.97\,km to $\sim$30\,m.

	\begin{deluxetable*}{clccc}
		\tablecaption{Alignment Attempt Summary \label{tab:alignment_attempts}}
		\tablehead{
			\colhead{\#} &
			\colhead{Method} &
			\colhead{Reference} &
			\colhead{Vikram Position} &
			\colhead{Offset from ISRO}
		}
		\startdata
		0 & Unaligned (raw SPICE) & --- & $-$69.177$^\circ$, 32.280$^\circ$ & 5.97\,km \\
		1 & \texttt{pc\_align} rigid ICP & LOLA 20\,m & $-$69.117$^\circ$, 32.325$^\circ$ & 7.78\,km \\
		2 & \texttt{pc\_align} translation-only & LROC NAC 3\,m & $-$69.348$^\circ$, 32.325$^\circ$ & $\sim$750\,m \\
		3 & Manual 5-crater tie-points & LROC NAC 3\,m & $\sim$$-$69.373$^\circ$, $\sim$32.319$^\circ$ & $\sim$30\,m \\
		\enddata
		\tablecomments{Reference Vikram coordinates: ISRO official $-$69.3733$^\circ$S, 32.3186$^\circ$E \citep{ISRO2023}; independently identified at $-$69.3741$^\circ$S, 32.32$^\circ$E by \citet{LROC2023}. Row~1 is included to illustrate why LOLA alignment failed at this site: the ICP solution optimized $Z$ but moved the DEM \emph{away} from the correct $XY$ position. Row~3 is the final product.}
	\end{deluxetable*}

	\begin{figure*}
		\centering
		\includegraphics[width=0.32\textwidth]{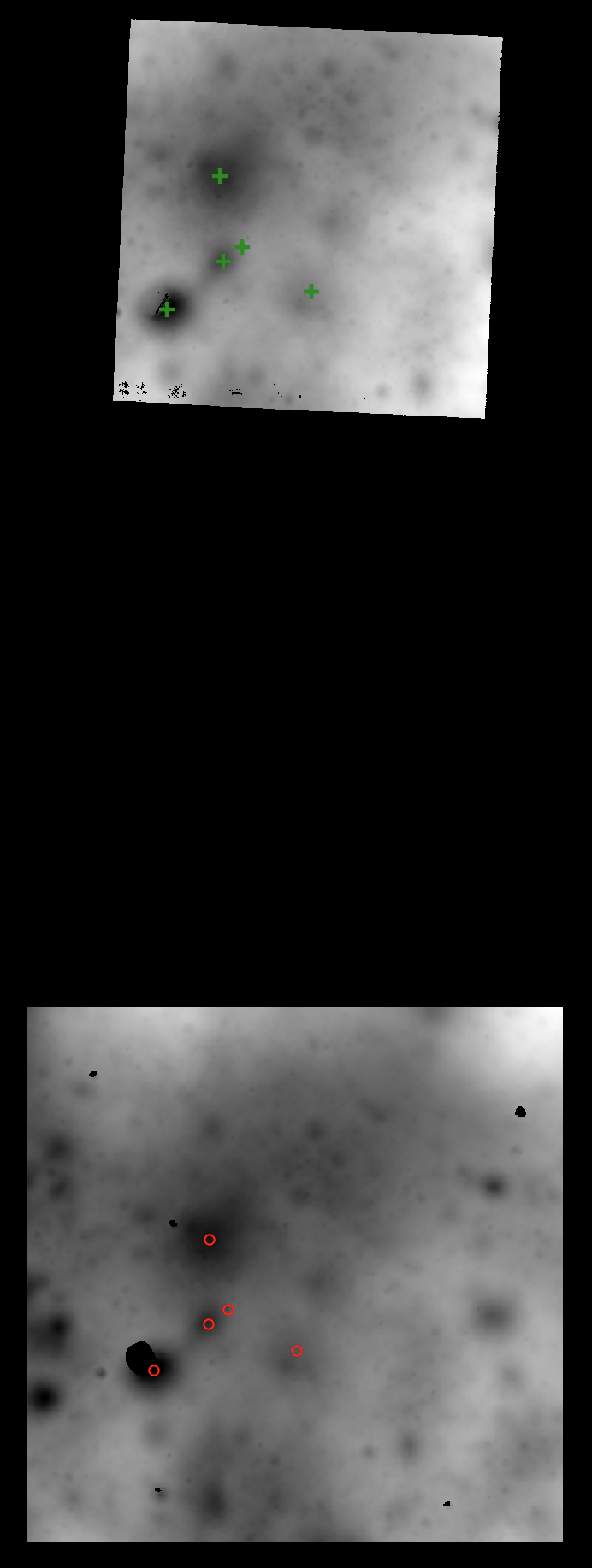}\hfill
		\includegraphics[width=0.32\textwidth]{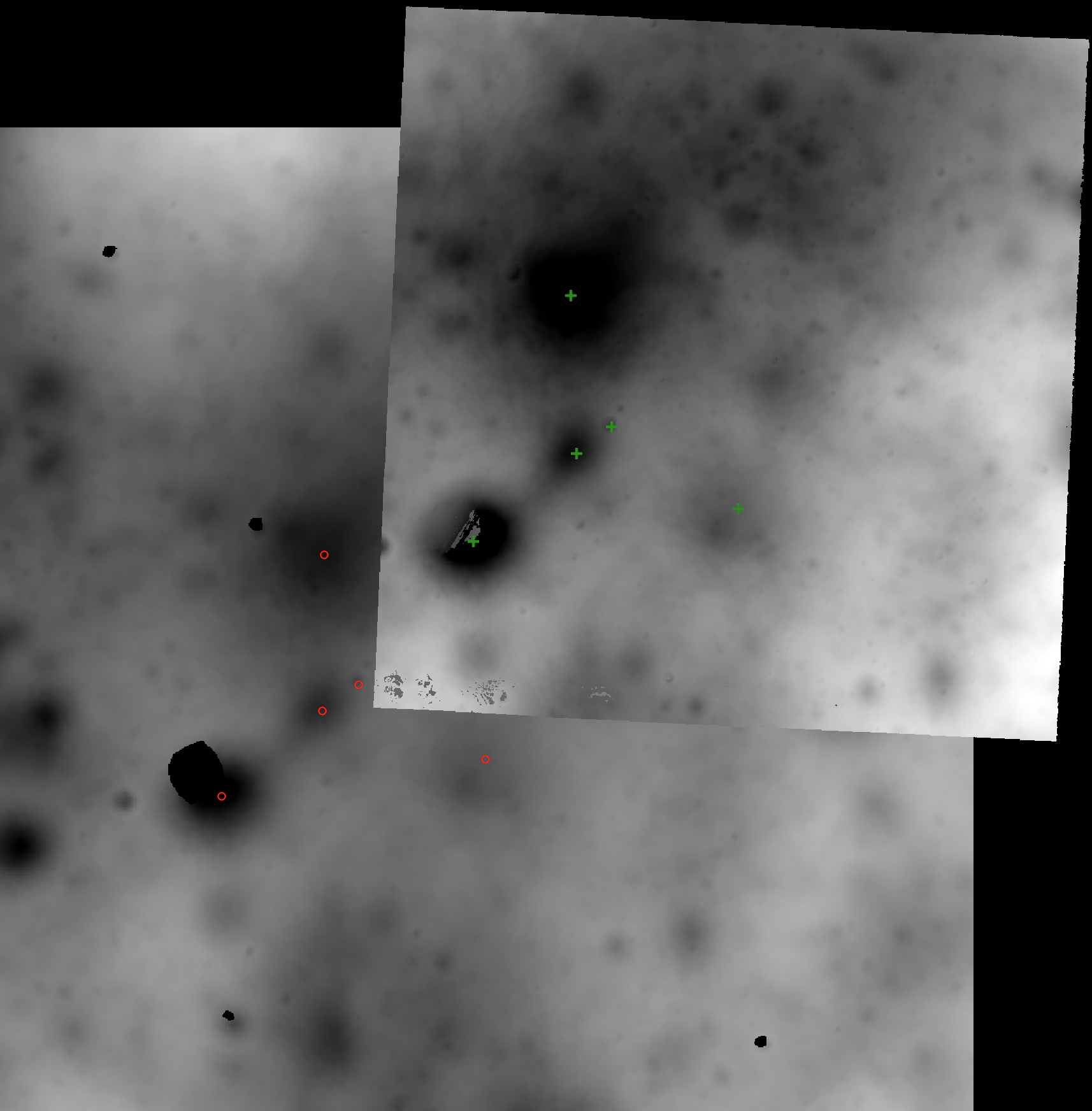}\hfill
		\includegraphics[width=0.32\textwidth]{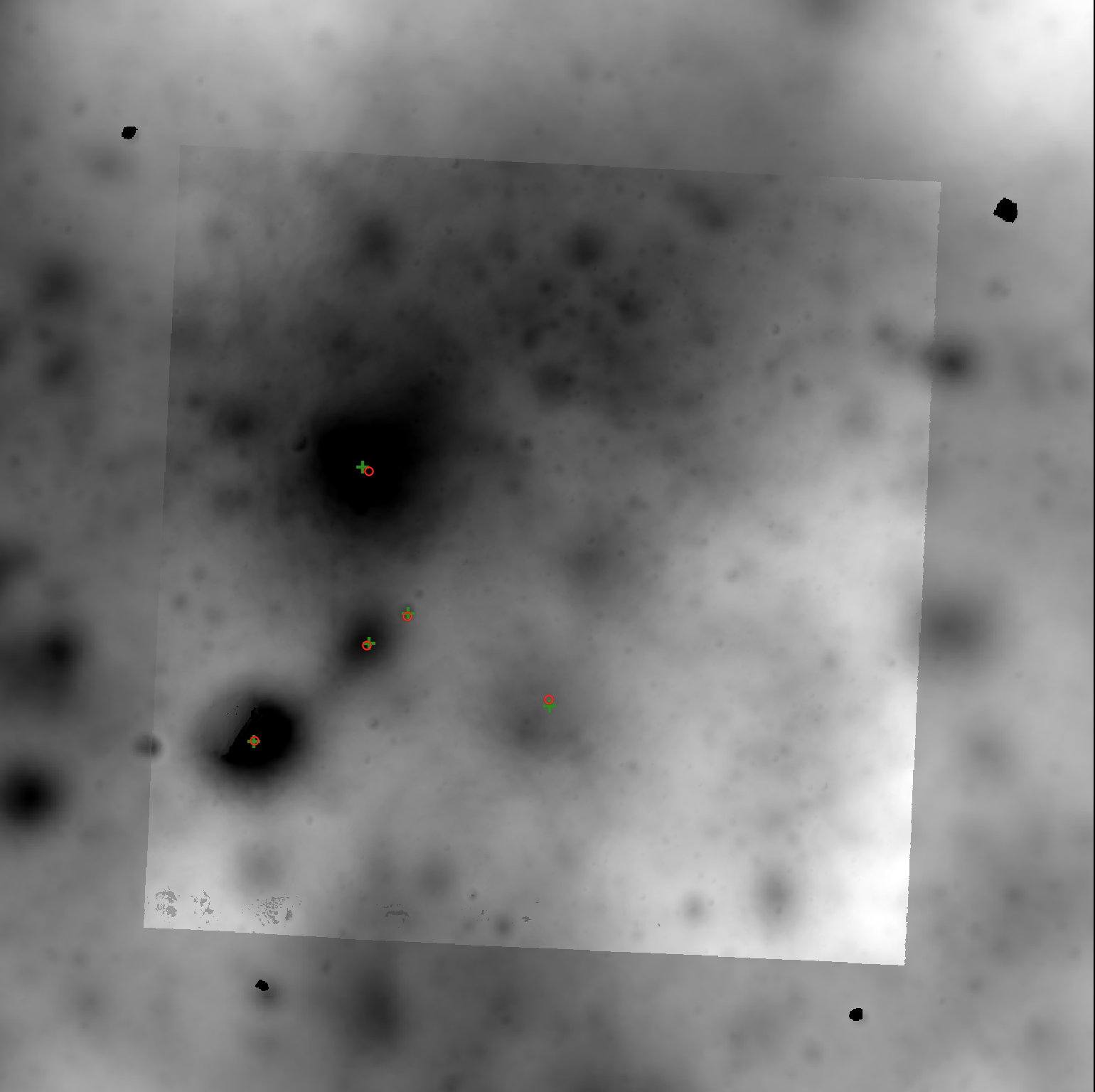}
		\caption{Progressive alignment of the OHRC DEM (green crosses) to the LROC NAC reference surface (red circles), with tie-point crater locations marked in both datasets. \textit{Left:} Before alignment, showing the $\sim$5.97\,km systematic offset --- the OHRC and LROC NAC DEMs do not overlap. \textit{Center:} After automated \texttt{pc\_align} translation-only alignment, reducing the offset to $\sim$750\,m --- the DEMs overlap but crater markers remain visibly displaced. \textit{Right:} After manual five-crater tie-point refinement, achieving $\sim$30\,m registration accuracy --- corresponding markers coincide.}
		\label{fig:alignment_stages}
	\end{figure*}

	\begin{figure*}
		\centering
		\includegraphics[width=\textwidth]{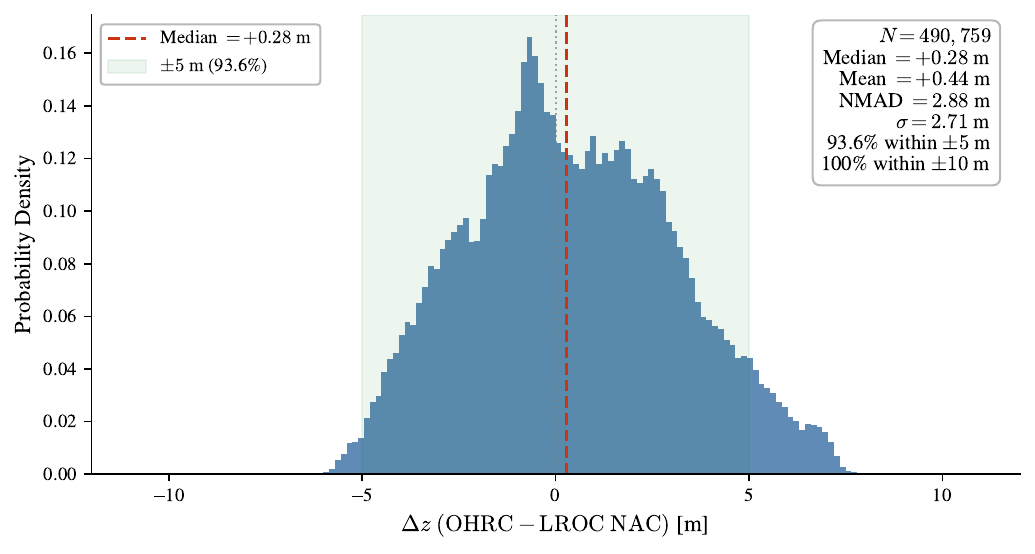}
		\caption{Distribution of elevation residuals ($\Delta z$ = OHRC $-$ LROC NAC) at 3\,m resolution after tie-point alignment ($N$ = 490{,}759). The median residual is $+$0.28\,m; the NMAD is 2.88\,m. 93.6\% of residuals fall within $\pm$5\,m.}
		\label{fig:residual_hist}
	\end{figure*}

	\subsection{Propagation to Full-Resolution DEM}

	The tie-point-derived transform was applied to the full-resolution stereo point cloud using \texttt{pc\_align} with zero additional iterations (\texttt{-{}-num-iterations 0}), propagating the translation without further optimization.
	The transformed point cloud was then rasterized at native resolution using \texttt{point2dem}.
	Table~\ref{tab:pinned_dem} summarizes the final product properties.

	\begin{deluxetable}{lc}
		\tablecaption{Final Pinned DEM Properties \label{tab:pinned_dem}}
		\tablehead{\colhead{Parameter} & \colhead{Value}}
		\startdata
		Resolution & 0.303\,m \\
		Valid pixels & 91.16\% \\
		Projection & Oblique stereographic (D\_MOON) \\
		Vertical reference & D\_MOON datum \\
		Alignment method & LROC NAC tie-point registration \\
		\enddata
	\end{deluxetable}

	\subsection{Reprojection and Validation Grid Preparation}
	\label{sec:reproj}

	To enable pixel-wise comparison between the OHRC and LROC NAC DEMs, the aligned OHRC DEM was reprojected from its native oblique stereographic CRS (centered on $-$69.1772$^\circ$, 32.3255$^\circ$) to the LROC NAC polar stereographic CRS (latitude of origin $-$90$^\circ$).
	Both DEMs were then resampled to a common 3\,m grid using bilinear interpolation, with identical spatial extent, pixel spacing, and nodata definitions to ensure strict pixel-to-pixel correspondence for the validation analysis described in Section~\ref{sec:vertical_validation}.

	\subsection{Vertical Accuracy Validation}
	\label{sec:vertical_validation}

	\subsubsection{Elevation Difference Raster}

	Following alignment and reprojection to a common 3\,m grid (Section~\ref{sec:reproj}), a pixel-wise elevation difference raster was computed as:
	\begin{equation}
		\Delta z = z_{\mathrm{OHRC}} - z_{\mathrm{NAC}}
	\end{equation}
	where $z_{\mathrm{OHRC}}$ is the aligned OHRC DEM elevation and $z_{\mathrm{NAC}}$ is the LROC NAC reference elevation, both on the common 3\,m grid.
	Positive $\Delta z$ indicates OHRC elevations higher than the LROC NAC surface.
	Only pixels valid in both DEMs were included; a total of $N$ = 490{,}759 valid pixel comparisons were obtained --- nearly two orders of magnitude more than the 5{,}076 pixels available in the earlier LOLA-based comparison at 20\,m.
	All validation statistics were derived directly from this $\Delta z$ raster.

	Figure~\ref{fig:dz_map} shows the spatial distribution of elevation residuals.
	The map reveals no coherent regional gradient or planar tilt across the overlap area.
	Residual variations are spatially localized and bounded, consistent with terrain-driven differences between the two independent stereo reconstructions rather than systematic geodetic misalignment.

	\begin{figure}
		\centering
		\includegraphics[width=\columnwidth]{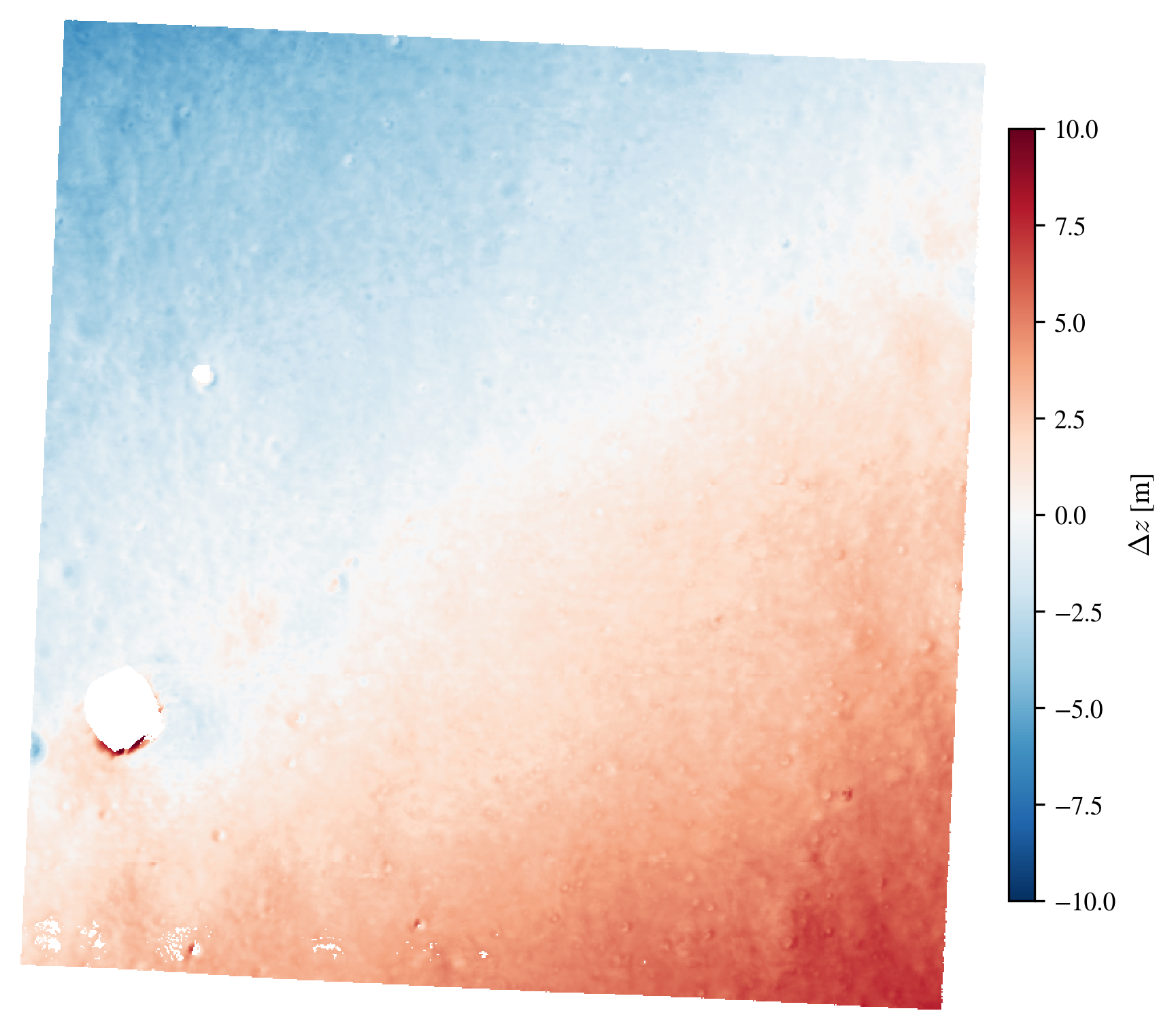}
		\caption{Elevation difference map ($\Delta z$ = OHRC $-$ LROC NAC) at 3\,m resolution after tie-point alignment and reprojection to a common grid. The absence of large-scale gradients indicates negligible residual tilt or systematic distortion. Most residuals fall within $\pm$5\,m, consistent with the reported NMAD of 2.88\,m.}
		\label{fig:dz_map}
	\end{figure}

	\subsubsection{Summary Statistics}

	The spatial distribution shown in Figure~\ref{fig:dz_map} is quantified below using robust statistical metrics derived from the $\Delta z$ raster \citep{Hohle2009}.
	In particular, the normalized median absolute deviation (NMAD $= 1.4826 \times \mathrm{median}(|\Delta z - \mathrm{median}(\Delta z)|)$) provides an outlier-resistant measure of dispersion.
	Table~\ref{tab:dz_stats} summarizes the descriptive statistics.

	\begin{deluxetable}{lc}
		\tablecaption{Descriptive Statistics of Elevation Residuals (OHRC $-$ LROC NAC, 3\,m grid) \label{tab:dz_stats}}
		\tablehead{\colhead{Metric} & \colhead{Value (m)}}
		\startdata
		Mean & $+$0.44 \\
		Median & $+$0.28 \\
		Standard deviation ($\sigma$) & 2.71 \\
		NMAD & 2.88 \\
		Valid pixels ($N$) & 490{,}759 \\
		\enddata
	\end{deluxetable}

	The median residual of $+$0.28\,m indicates negligible vertical bias following tie-point alignment.
	The close agreement between the mean ($+$0.44\,m) and median confirms a near-symmetric residual distribution.
	The robust dispersion (NMAD = 2.88\,m) is slightly larger than the standard deviation (2.71\,m), indicating mild heavy-tail behavior consistent with localized terrain differences at crater edges.

	\subsubsection{Percentile and Threshold Metrics}

	Tables~\ref{tab:dz_percentiles} and~\ref{tab:dz_thresholds} provide percentile-based and threshold-based accuracy assessments.

	\begin{deluxetable}{lc}
		\tablecaption{Percentile Statistics of Elevation Residuals \label{tab:dz_percentiles}}
		\tablehead{\colhead{Percentile} & \colhead{$|\Delta z|$ (m)}}
		\startdata
		P90 & 4.49 \\
		P95 & 5.26 \\
		P99 & 6.69 \\
		\enddata
	\end{deluxetable}

	\begin{deluxetable}{lc}
		\tablecaption{Fraction of Residuals Within Vertical Thresholds \label{tab:dz_thresholds}}
		\tablehead{\colhead{Threshold} & \colhead{Fraction}}
		\startdata
		$|\Delta z| \leq 5$\,m & 93.6\% \\
		$|\Delta z| \leq 10$\,m & 100.0\% \\
		\enddata
	\end{deluxetable}

	93.6\% of residuals fall within $\pm$5\,m and 100\% within $\pm$10\,m, representing a substantial improvement over the earlier LOLA-based validation (75.8\% within $\pm$10\,m at 20\,m resolution).

	\subsubsection{Spatial Characteristics}

	The $\Delta z$ raster shows no coherent regional gradient, tilt, or systematic warping, indicating successful geodetic registration.
	The observed dispersion is interpreted as arising from differences between the two independent stereo reconstruction methods, terrain slope amplification at crater edges, and residual stereo reconstruction noise in both products.

	\subsubsection{Validation Limitations}

	This validation constrains accuracy at the 3\,m grid resolution and does not directly assess sub-grid microtopographic accuracy or performance on extremely steep slopes where interpolation effects may amplify residuals \citep{Barker2021,Mazarico2018}.
	The LROC NAC DEM itself has a relative vertical precision of $\sim$1.9\,m \citep{Henriksen2016}; the reported NMAD therefore reflects the combined uncertainty from both the OHRC DEM and the LROC NAC reference surface.
	Compared to the previous LOLA-based validation, the LROC NAC comparison provides nearly 100$\times$ more valid pixels, 3$\times$ finer spatial resolution, and substantially tighter dispersion metrics, yielding a more definitive assessment of the OHRC DEM's vertical accuracy.

	\subsubsection{Summary}

	The aligned OHRC DEM exhibits negligible vertical bias ($+$0.28\,m) and a robust vertical dispersion of 2.88\,m (NMAD) relative to the LROC NAC DEM at 3\,m scale.
	The residual distribution is symmetric, tightly bounded ($\pm$10\,m), and free of systematic distortion, confirming effective geodetic alignment and reliable absolute vertical control.
	The 3\,m-scale validation, with nearly 500{,}000 pixel comparisons, provides strong confidence in the accuracy of the full-resolution DEM for geomorphological analysis and slope characterization.

	\section{Discussion}
	\label{sec:discussion}

	\subsection{Significance and Camera Model Selection}

	This work demonstrates that the OHRC archive on PRADAN contains sufficient information --- calibrated imagery, SPICE kernels, and metadata --- for researchers to produce science-grade elevation products without access to ISRO's proprietary OPTIMUS software, validating the open-source planetary photogrammetry ecosystem (ISIS, ASP, ALE) for Chandrayaan-2 data processing. The methodology is directly applicable to other OHRC stereo pairs, including the extensive polar imaging campaign conducted in early 2024 in preparation for the LUPEX mission.

	The performance difference between ISIS and CSM camera models (Section~\ref{sec:csm_comparison}) has practical implications beyond this study. CSM-based models were necessary to achieve viable stereo convergence in our experiments, and this finding may be relevant to other pushbroom instruments with similar stereo acquisition geometries. We recommend that it be documented in the ASP user guide for Chandrayaan-2 workflows.

	\subsection{Comparison with Existing Products}

	Table~\ref{tab:comparison} compares the present DEM with existing lunar elevation datasets, including the experimental result from ASP's own Chandrayaan-2 documentation.

	\begin{deluxetable*}{lccc}
		\tablecaption{Comparison with Existing Lunar Elevation Products \label{tab:comparison}}
		\tablehead{
			\colhead{Product} &
			\colhead{Horizontal Resolution} &
			\colhead{Vertical Precision} &
			\colhead{Source}
		}
		\startdata
		LOLA global grid & $\sim$30\,m & $\sim$1\,m & NASA/GSFC \\
		LOLA high-density tracks & $\sim$1\,m (along-track) & $\sim$10\,cm & NASA/GSFC \\
		LROC NAC DEM & $\sim$2\,m & $\sim$0.5--1\,m & NASA/GSFC/ASU \\
		OHRC DEM (OPTIMUS, SAC) & $\sim$0.28\,m & $\sim$1\,m & ISRO SAC \\
		OHRC DEM (ASP example)\tablenotemark{a} & 1\,m & $\sim$0.25--0.50\,m\tablenotemark{b} & NASA Ames \\
		OHRC DEM (this work) & $\sim$0.30\,m & $\sim$0.40--0.50\,m & Independent \\
		\enddata
		\tablenotetext{a}{Experimental workflow described in ASP documentation.}
		\tablenotetext{b}{Reported jitter artifacts on the order of the image GSD ($\sim$0.25\,m); required manual LOLA alignment after $\sim$4\,km initial offset.}
	\end{deluxetable*}

	\subsection{Applications}

	DEMs at the 30\,cm scale open applications not feasible with coarser products, including sub-meter hazard identification (boulders, small craters, slopes) for upcoming missions such as Chandrayaan-4, LUPEX, and Artemis; post-landing forensic analysis of surface disturbances and hardware orientations; regolith texture characterization at decimeter scales; and slope mapping for rover traverse planning.

	\subsection{Validation and Limitations}

	Direct validation of the DEM against an independent reference at comparable resolution is not currently possible: ISRO's OPTIMUS-derived OHRC DEMs have not been publicly released, and no other sub-meter elevation product exists for this region.
	We therefore rely on a combination of indirect assessments.
	The reported triangulation error and slope statistics quantify relative geometric precision and do not constitute absolute vertical accuracy with respect to a lunar geodetic datum.

	\begin{enumerate}
		\item \textbf{Internal consistency:} The median triangulation error of 8.1\,cm and valid pixel fraction of 91.2\% indicate robust stereo reconstruction. The $\sim$94{,}000 feature matches across the stereo pair confirm dense, reliable correspondence.
		\item \textbf{Triangulation error map characterization:} The spatial distribution of the triangulation error (Figure~\ref{fig:error}) shows that the majority of the coverage area exhibits errors well below the median, with elevated errors confined to crater walls (where occlusion and shadow degrade stereo matching), strip margins (where stereo geometry is weakest), and along-track bands correlated with ASP's tile-based processing boundaries. The Chandrayaan-3 Vikram landing site lies within a predominantly low-error zone. An along-track banding pattern is visible, attributable to the tile blending step in ASP's MGM stereo algorithm --- this is a known characteristic of the current implementation and does not indicate a failure in the stereo geometry.
		\item \textbf{Terrain characterization consistency:} \citet{Amitabh2023} characterized the terrain at Chandrayaan-3 candidate landing sites (designated S1--S7) using OPTIMUS-derived DEMs, identifying the selected site for its favorable slope and roughness properties. Our DEM of the same landing site shows widely spaced elevation contours and low local slopes, consistent with these reported terrain characteristics.
		\item \textbf{Geodetic alignment and vertical validation:} The initial DEM exhibited a $\sim$5.97\,km systematic offset. An initial LOLA-based alignment failed to constrain horizontal position at this featureless site (Section~\ref{sec:lola_failed}). Subsequent alignment to an LROC NAC DEM using automated registration and manual five-crater tie-points (Section~\ref{sec:nac_alignment}) reduced the offset to $\sim$30\,m. Pixel-wise comparison with the LROC NAC surface at 3\,m resolution (Section~\ref{sec:vertical_validation}) yields a median residual of $+$0.28\,m and NMAD of 2.88\,m, with 93.6\% of residuals within $\pm$5\,m, confirming negligible vertical bias and reliable absolute elevation control.
		\item \textbf{Feature verification:} The Chandrayaan-3 Vikram lander is resolved as a distinct elevation feature, confirming the effective spatial resolution of the product. Following geodetic alignment, the lander position in the DEM is consistent with published landing coordinates.
	\end{enumerate}

	\subsubsection{Cross-Validation with LROC NAC}
	\label{sec:planned_validation}

	Cross-validation against the LROC NAC DEM at 3\,m resolution (Section~\ref{sec:vertical_validation}) provides a substantially more rigorous assessment than LOLA-based validation alone.
	The LROC NAC comparison yields nearly 500{,}000 valid pixel comparisons (vs.\ $\sim$5{,}000 for LOLA at 20\,m), median bias of $+$0.28\,m (vs.\ $-$1.47\,m), and NMAD of 2.88\,m (vs.\ 9.11\,m).
	The threefold improvement in dispersion and the elimination of vertical bias confirm both the accuracy of the tie-point alignment and the geometric fidelity of the OHRC stereo reconstruction at meter scales.

	\subsubsection{Known Artifacts and Future Improvements}

	The ASP Chandrayaan-2 documentation notes jitter artifacts on the order of the image GSD ($\sim$0.25\,m) and unmodeled lens distortion in OHRC stereo products.
	These artifacts are also present to some degree in our DEM, manifesting as the along-track banding visible in the triangulation error map.
	Importantly, these are not fundamental limitations of the OHRC instrument or the data --- they reflect the current state of the open-source tool chain, which is under active development:

	\begin{itemize}
		\item ASP includes a jitter correction tool (\texttt{jitter\_solve}, Section 16.39 of the ASP documentation) that models and removes high-frequency attitude oscillations. Applying this to OHRC data is expected to reduce the banding pattern and improve vertical accuracy.
		\item ALE pull request \#682 introduces geometric corrections to the OHRC camera driver that may reduce lens distortion residuals.
		\item Future ASP releases may incorporate improved tile blending strategies for the MGM stereo algorithm, further reducing along-track artifacts.
	\end{itemize}

	As these tools mature, reprocessing the same stereo pair should yield measurably improved results without any change to the input data or the overall methodology.

	Several additional caveats apply:

	\begin{itemize}
		\item The vertical precision estimate (40--50\,cm) is derived from the triangulation error distribution and the stereo geometry; it has not been verified against an external reference and should be regarded as relative precision rather than absolute accuracy.
		\item Platform-specific configuration details are documented in the supplementary notes \citep{\suppref\suppyear}.
	\end{itemize}

	The OHRC archive contains stereo pairs at multiple locations including the lunar south pole, mid-latitude geological sites, and landing sites of other missions.
	We plan to process additional sites using the methodology described here.
	The Soma portal (Section~\ref{sec:data}) will continue to serve as the primary tool for stereo pair discovery across the OHRC catalogue.

	\section{Conclusions}
	\label{sec:conclusions}

	We present a 0.30\,m\,pixel$^{-1}$ digital elevation model of the Chandrayaan-3 Vikram landing site derived from Chandrayaan-2 OHRC stereo imagery using a fully open photogrammetric workflow based on ISIS, the Ames Stereo Pipeline, and ALE.
	The reconstruction achieves a median triangulation error of 8.1\,cm with 91.2\% valid pixel coverage across a 2.18\,$\times$\,2.24\,km region, corresponding to an estimated relative vertical precision of 40--50\,cm.
	At sub-meter scale, the Vikram lander and Pragyan rover are individually resolved as discrete topographic features, demonstrating the geometric fidelity achievable from OHRC stereo data.
	Cross-validation against the LROC NAC DEM at 3\,m resolution confirms negligible vertical bias (median $\Delta z$ = $+$0.28\,m) and robust dispersion (NMAD = 2.88\,m), with no coherent regional gradients in the residual surface.

	A central technical result is the demonstrated necessity of Community Sensor Model (CSM) camera models for stable OHRC pushbroom stereo reconstruction: the legacy ISIS camera model failed to produce stable solutions under identical conditions across two independent sites.
	This finding, together with the geodetic alignment strategy developed to overcome the $\sim$6\,km initial offset inherent to OHRC pitch-maneuver stereo, provides a reproducible template for processing additional stereo pairs from the OHRC archive.

	The practical value of sub-meter OHRC topography has already been demonstrated operationally: ISRO's terrain characterization of Chandrayaan-3 candidate landing sites \citep{Amitabh2023,DurgaPrasad2024} and the vision-based hazard detection and avoidance system that guided Vikram to the first successful south polar landing \citep{Suresh2024} relied on OHRC-derived DEMs produced with the proprietary OPTIMUS pipeline.
	The open workflow presented here yields comparable spatial resolution and vertical precision, providing the broader community with an independent capability to produce equivalent products from publicly archived imagery.
	At 0.30\,m, these DEMs complement the LROC NAC DTMs ($\sim$1\,m resolution) that currently serve as the standard reference for lunar surface operations, resolving sub-meter hazards --- boulders, small craters, steep local slopes --- that fall below the NAC detection threshold.

	Applied to the extensive OHRC south polar archive, this methodology enables independent hazard mapping, slope analysis, and terrain characterization for upcoming missions including Chandrayaan-4, LUPEX, and Artemis.
	The workflow executes in approximately 19 minutes on a single workstation, and the Soma portal provides stereo pair discovery across the full OHRC catalogue.
	The resulting DEM, orthoimage, and triangulation error products will be archived with a DOI to support independent reuse.

	\section*{Data Availability}

	The derived digital elevation model (DEM), orthorectified imagery, and supporting products are archived with a dedicated DOI upon acceptance.
	\ifdeanon{%
		Supplementary technical documentation describing the OHRC stereo processing workflow and configuration settings is archived at Zenodo: \url{https://doi.org/10.5281/zenodo.18634148} \citep{Tungathurthi2026}.
	}{%
		Supplementary technical documentation describing the OHRC stereo processing workflow and configuration settings is archived at Zenodo (DOI withheld for review) \citep{Anonymous2026}.
	}
	The source Chandrayaan-2 OHRC imagery and SPICE kernels are publicly available through ISRO's PRADAN portal (\url{https://pradan.issdc.gov.in/ch2/}).
	All tools used in the pipeline (ISIS, ASP, ALE) are open-source.
	All imagery derived from Chandrayaan-2 data is \textcopyright\ ISRO.

	\begin{acknowledgments}
		This work was made possible by the open-source planetary science tools developed and maintained by the USGS Astrogeology Science Center (ISIS, ALE), NASA Ames Research Center Intelligent Robotics Group (Ames Stereo Pipeline), and the broader Community Sensor Model consortium.

		We acknowledge the use of data from the Chandrayaan-II, second lunar mission of the Indian Space Research Organisation (ISRO), archived at the Indian Space Science Data Centre (ISSDC).
		All Chandrayaan-2 data remain the property of ISRO; \textcopyright\ reserved ISRO.

		The author acknowledges the OHRC instrument team at ISRO's Space Applications Centre for building and operating the instrument.
	\end{acknowledgments}

	\software{ISIS \citep{ISIS2023}, ASP \citep{Beyer2018}, ALE \citep{ALE2023}, USGSCSM, GDAL, QGIS, Qgis2threejs \citep{qgis2threejs}, matplotlib, Python}

	\facilities{Chandrayaan-2 (OHRC), LRO (LOLA, LROC NAC)}

	\bibliographystyle{aasjournalv7}
	\bibliography{references}

\end{document}